\documentclass{aastex63}

\newcommand\rmi{\mathrm{i}}
\newcommand{\comment}[1]{}

\usepackage{graphicx,subfigure}

\begin{document}

\title{The Drag Instability in a 2D Isothermal C-shock}

\correspondingauthor{Pin-Gao Gu}
\email{gu@asiaa.sinica.edu.tw}

\author{Pin-Gao Gu}
\affiliation{Institute of Astronomy \& Astrophysics, Academia Sinica,
Taipei 10617, Taiwan}

\begin{abstract}
We extend the linear analysis of the drag instability in a 1D perpendicular isothermal C-shock by Gu \& Chen to 2D perpendicular and oblique C-shocks in the typical environment of star-forming clouds. Simplified dispersion relations are derived for the unstable modes. We find that the mode property of the drag instability generally depends on the ratio of the transverse (normal to the shock flow) to longitudinal (along the shock flow) wavenumber. For the  transversely large-scale mode,  the growth rate and wave frequency of the drag instability in a 2D shock resemble those in a 1D shock. For the transversely small-scale mode, the drag instability is characterized by an unstable mode coupled with an acoustic mode primarily along the transverse direction. When the shock is perpendicular or less oblique, there exists a slowly propagating mode, which can potentially grow into a nonlinear regime and contribute to the maximum growth of the instability.  In contrast, when the shock is more oblique, this slowly propagating unstable mode disappears, and the maximum growth of the drag instability is likely contributed from the transversely large-scale mode (i.e., almost 1D mode). In all cases that we consider, the magnitude of the density perturbations is significantly larger than that of the velocity and magnetic field perturbations,  implying that the density enhancement governs the dynamics in the linear regime of the instability. A few issues in the linear analysis, as well as the possible astrophysical implications, are also briefly discussed.
\end{abstract}

\section{Introduction}
\label{sec:intro}
Stars form in the cold and dense cores of molecular clouds \citep[e.g.,][]{KE12,HI19,Girichidis20}. Many observations suggest that molecular clouds exhibit highly supersonic turbulence \citep[e.g.,][]{ES04,BP07,HF12}. Additionally, cosmic rays can weakly ionize the neutral gas to produce ions with a typical ionization fraction $\lesssim 10^{-6}$ \citep[e.g.,][]{Draine,Tielens05,Dalgarno06,Indriolo12}. Since the magnetic fields are normally assumed to be tightly coupled to the ions, which in turn interact with the neutrals through the drag force, the ions together with the magnetic fields can systematically drift relative to the neutrals, a phenomenon known as ambipolar diffusion \citep{Spitzer1956,MS1956,Shu,Zweibel}. 

Of particular interest in this study is the interplay between ambipolar diffusion and shocks.  While jump-type (J-type) shocks exhibit sharp supersonic (or super-Alfv\'enic for magnetized shocks) discontinuities in physical properties, in the case of nonideal magnetohydrodynamics (MHD), the different signal speeds of the ions and neutrals along with the ambipolar diffusion broaden the discontinuity, leading to continuous-type (C-type) shocks with a smooth transition in physical quantities between the pre- and post-shock regions \citep{Draine,DM93}. 
Observationally, various efforts have been made
 to detect such features in turbulent molecular clouds \citep[e.g.,][]{LiHoude08,Hezareh10,Hezareh14,XuLi16,Tang18}, though most of them are indirect measurements and highly dependent on the adopted dynamical and chemical models \citep[e.g.,][see also the Introduction of Gu \& Chen 2020, hereafter \citet{GC20}]{Flower98,Flower10,Gusdorf08,LehmannWardle16,Valdivia17}.
Theoretically, 
\citet{CO12,CO14} analyzed the structures of perpendicular and oblique C-shocks in exquisite detail in the typical environment of star-forming clouds and extended their work to colliding flows to explore the formation of cores and filaments in numerical simulations.

However, a C-shock is not necessarily a dynamically stable structure. Based on the 1D background state of steady, plane-parallel C-shocks derived by \citet{CO12},  \citet{GC20} conducted a Wentzel-Kramers-Brillouin-Jeffreys (WKBJ) analysis and confirmed the postulation of \citet{Gu04} that the drag instability can occur in a 1D isothermal perpendicular C-shock. In an environment where the ambipolar diffusion is efficient and the ionization--recombination equilibrium is nearly attained, the drag instability, a plasma effect discovered by \citet{Gu04},  ensues from the ion-neutral drag and is a local linear overstability phenomenon associated with an exponentially growing mode of a propagating wave. On the other hand, it is well known that C-shocks are also susceptible to the Wardle instability \citep{Wardle,Wardle1991}. However, the Wardle instability is suppressed by the fast ionization--recombination process expected to pervade star-forming clouds \citep[e.g.,][]{SmithML97,Stone,Falle}.

It was found by \citet{GC20} that the growing wave mode of the drag instability can only propagate downstream within a 1D shock and subsequently decay in the post-shock region. The authors demonstrated that the growth of the drag instability in a 1D perpendicular shock is limited by the short time span for the unstable wave to stay within a C-shock. Consequently, the maximum growth of an unstable wave is given by the maximum total growth (MTG), which is defined by \citet{GC20} as the maximum value of the total growth
of an unstable mode traveling over an entire shock width before it is damped in the post-shock region. Specifically, MTG$=\exp\int_{\rm shock\ width} \Gamma_{grow} dx/v_{ph,x}$, where $\Gamma_{grow}$ is the growth rate and $v_{ph,x}$ is the longitudinal phase velocity. 
The authors estimated the MTG for such an unstable wave in typical environments of star-forming clouds.
They found that a stronger shock with a larger shock width favors a more appreciable growth of the 1D drag instability. Most importantly, the linear analysis suggests that the density enhancement induced by the drag instability dominates over the magnetic field and velocity enhancements in the dynamical growth of the unstable mode.

One of the most important questions for shocks in this context is whether a shock instability can lead to fragmentation that subsequently undergoes gravitational collapse and eventually induces star formation. The predominant density enhancement of the drag instability provides a possible mechanism to form supercritical clumps/cores within a C-shock in the typical environments of star-forming clouds. As a preliminary activity, the analysis by \citet{GC20} focused on the basic behavior of the drag instability within the steady-state profiles of C-shocks, which is 1D, linear, and non-self-gravitating. Hence, how exactly this plasma instability plays 
a role in the observational evidence of prestellar cores and filamentary networks 
in relation to magnetic field morphology \citep[e.g.,][]{Andre14,LiHB14} is not clear and has yet to be addressed in terms of the current theoretical framework in 1D. Obviously, the theory is still in its infancy and requires further development toward more realistic physical applications. 

Therefore, as a natural consequence of an ongoing effort, we extend the 1D analysis of the drag instability by \citet{GC20}  and conduct a 2D linear analysis for both perpendicular and oblique shocks in this study.  The aim is to develop a more comprehensive analytical work to provide useful information for probing and characterizing the drag instability in future numerical simulations and proceed to a more realistic case for future astrophysical applications.

The contents of the paper are structured as follows. In Section~\ref{sec:perp}, we begin with a linear analysis of the drag instability in a 2D perpendicular shock and identify the mode with the maximal growth in the fiducial model of a C-shock. Based on that experience, we proceed to the linear analysis of the drag instability in an oblique shock in Section~\ref{sec:oblique} and study the instability properties in both the fiducial model and a model referred to as model V06 for a stronger and wider shock for comparison. 
Finally, the summary and brief discussions are presented in Section~\ref{sec:sum}.

\section{Linear analysis: isothermal perpendicular shocks}
\label{sec:perp}

In cold molecular clouds and their substructures, the dynamical evolution of ions and neutrals is governed by their individual continuity and momentum equations,
which include cosmic-ray ionization, ion--electron recombination in the gas phase, mutual collisional drag force, the Lorentz force on ions, and the pressure force with the isothermal equation of state.
Additionally, the evolution of magnetic fields is governed by the induction equation for ions. The entire set of equations then reads as follows \citep[e.g.,][]{Draine80,Shu,CO12,GC20}:
\begin{eqnarray}
{\partial \rho_n \over \partial t}+ \nabla \cdot (\rho_n {\bf v_n})=0, \label{eq:1}\\
{\partial \rho_i \over \partial t}+ \nabla \cdot (\rho_i {\bf v_i})= -\beta \rho_i^2 + \xi_\mathrm{CR} \rho_n, \label{eq:2} \\
\rho_n \left[ {\partial {\bf v_n} \over \partial t} + ({\bf v_n} \cdot \nabla ) {\bf v_n} \right]+\nabla p_n ={\bf f_d}, \label{eq:3}\\
\rho_i \left[ {\partial {\bf v_i} \over \partial t} + ({\bf v_i} \cdot \nabla ) {\bf v_i} \right]+\nabla p_i -{1\over 4\pi}(\nabla \times {\bf B})\times {\bf B}=-{\bf f_d}, \\
{\partial {\bf B} \over \partial t} + \nabla \times ({\bf B} \times {\bf v_i})=0, \label{eq:5}
\end{eqnarray}
where ${\bf v}$ is the velocity, $\rho$ is the density, ${\bf B}$ is the magnetic field, $p=\rho c_s^2$ is the gas pressure when the isothermal sound speed $c_s$ is 0.2~km/s at a temperature of $\sim 10$~K \citep{Fukui10}, and the subscripts $i$ and $n$ denote the ion and neutral species, respectively. 
Note that the neutrals and ions are coupled by the collisional drag force ${\bf f_d}\equiv \gamma \rho_i \rho_n {\bf v_d}=\gamma \rho_i \rho_n ({\bf v_i}-{\bf v_n})$, where $\gamma \approx 3.5 \times 10^{13}$~cm$^3$~s$^{-1}$~g$^{-1}$ is the drag force coefficient \citep{Draine}. 
The evolution of ion number density is controlled by the cosmic-ray ionization rate $\xi_\mathrm{CR}$ and the ion recombination in the gas phase $\beta$ \citep[see e.g.,][]{CO12}. We define the ionization parameter   $\chi_{i0} \equiv 10^6 \sqrt{\xi_\mathrm{CR} (m_n/m_i)/(\beta m_i)}$.  as done by \citet{CO12}.
We thus adopt $\beta \approx 10^{-7}$~cm$^3$~s$^{-1}/m_i$ and $\xi_\mathrm{CR} \approx 10^{-17}$~s$^{-1}$~($m_i/m_n$) in this study \citep[see, e.g.,][]{Shu,Tielens05}, where $m_n=2.3\times$ and $m_i=30 \times$ the hydrogen mass are considered.
Indeed, $\chi_{i0}=10$ falls in the typical range of $\chi_{i0}$ observed in star-forming regions ($\sim 1-20$; see, e.g.,~\citealt{McKee10}).

\subsection{Background States and Linearized Equations}

Since there is no background structure parallel to the shock front in a 2D plane-parallel, perpendicular shock, no fluid motion is normal to the shock flow. Resultantly, the background states of a 2D perpendicular shock are the same as those in a 1D perpendicular shock.
Therefore, a detailed description of the background states can be found in \citet{GC20}. For clarity, here we simply summarize the shock model to smoothly connect to the linear analysis to be presented later in the paper. We consider  the gas flow toward the $+x$ direction across the shock with the magnetic field in the $y$-direction. The equilibrium equations for a plane-parallel shock (i.e. $\partial/\partial t=\partial/\partial y=0$) are derived from Equations(\ref{eq:1})--(\ref{eq:5}),  which are given by Equations(6)--(10) in \citet{GC20} to provide the background state of our linear analysis in the shock frame.
In these equilibrium equations, \citet{GC20} also employed a strong coupling approximation under which the ion-neutral drag is balanced by the magnetic pressure gradient of the ions, i.e., $\gamma \rho_n L_B V_d=V_{A,i}^2$ where $L_B \equiv (-d \ln B/dx)^{-1}$ and $V_{A,i}$ is the Alfv\'en speed of the ions. Additionally, the equilibrium between cosmic ionization and recombination is assumed; namely, $\beta \rho_i^2 = \xi_\mathrm{CR} \rho_n$ (see \cite{CO12} for justifications of such a choice).

\begin{figure}
\plottwo{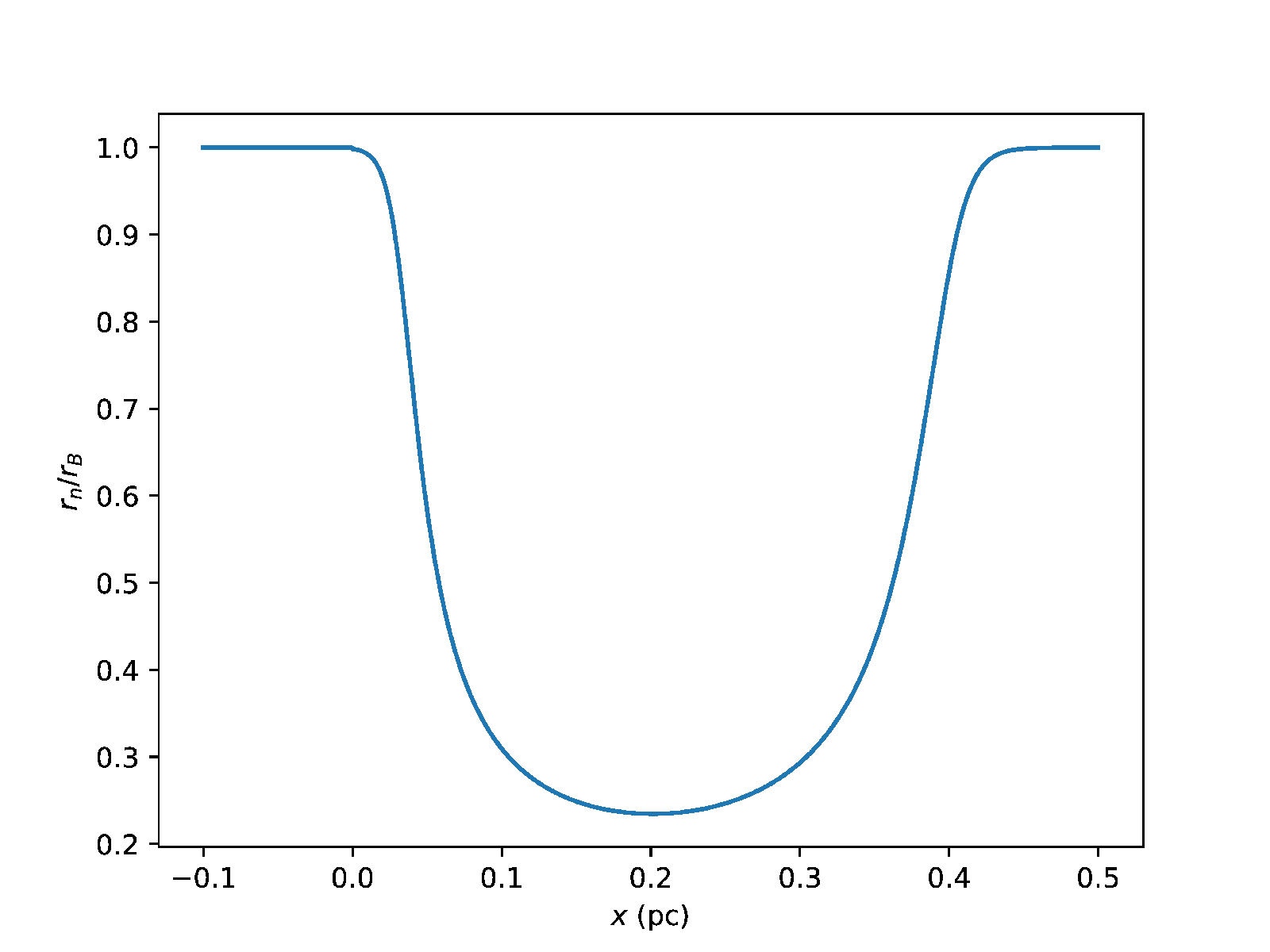}{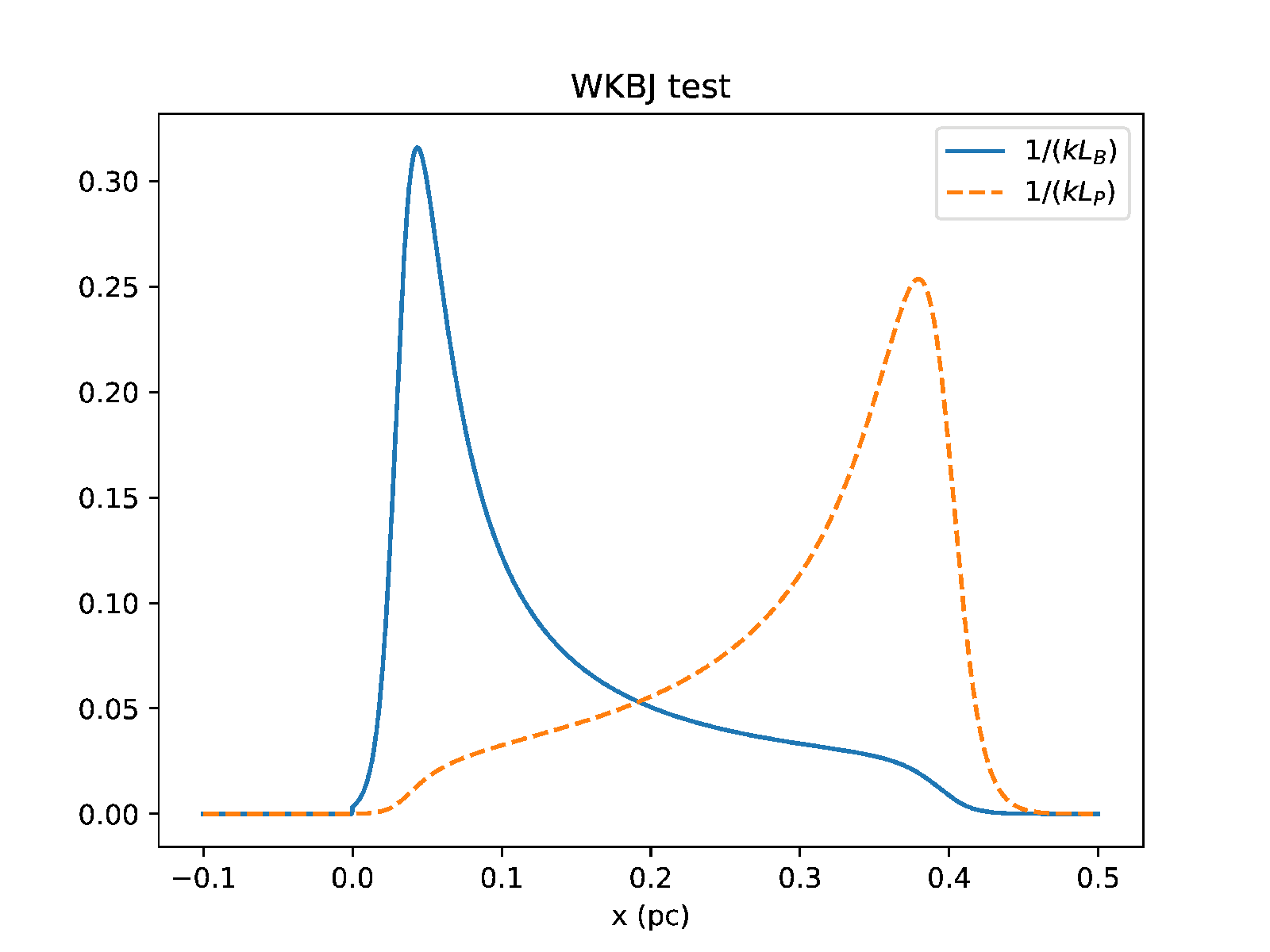}
\caption{The $r_n/r_B$ ratio of the background state throughout the C-shock in the fiducial model (left panel) and a test for the WKBJ approximation when $k_x=k_{fid}\equiv 1/0.015$ pc$^{-1}$ (right panel). Here $L_p \equiv |d\ln p/dx|^{-1}$ and $L_B $ are the scale heights for the gas pressure and magnetic field of the background states, respectively.}
\label{fig:fig1}
\end{figure}

Applying the zero-gradient boundary conditions ($d/dx=0$) far upstream and downstream (i.e., no structures in the steady pre- and post-shock regions), \citet{CO12} derived the 1D structure equation of a C-shock. Together with the pre-shock conditions described by $n_0$ (neutral number density), $v_0=V_{i,0}=V_{n,0}$ (shock velocity), $B_0$, and $\chi_{i0}$ (here and throughout this paper, we use the subscript 0 to denote a physical quantity in the pre-shock region),
the field compression ratio $r_B \equiv B/B_0 = V_{i,0}/V_i$ as well as
the neutral compression ratio $r_n \equiv \rho_n/\rho_{n,0}=V_{n,0}/V_n$ can be solved, and $\rho_i$, $\rho_n$, $B$, $V_i$, and $V_n$ can be subsequently obtained throughout the shock width. Following \citet{GC20}, we place the shock front at $x=0$ pc for convenience to plot and refer to shock properties as a function of $x$.
In this setup of the problem, the background drift velocity $V_d=V_i-V_n<0$ inside the C-shock (i.e. within the smooth shock transition).
As in \citet{GC20}, we adopt the 1D steady C-shock model shown in Figure 3 of \citet{CO12} as the fiducial model for
the background state in the shock frame, with the pre-shock parameters 
$n_0=500$ cm$^{-3}$, $v_0=5$ km/s, $B_0=5 \mu$G, and $\chi_{i0}=10$. The left panel of Figure~\ref{fig:fig1} shows $r_n/r_B$ in the fiducial model, which is the same as Figure 1 in \citet{GC20}.
%In this setup of the problem, the background drift velocity $V_d=V_i-V_n<0$ inside the C shock (i.e. within the smooth shock transition).

We now consider the perturbation eigenvector given by $U(\omega, k_x,k_y)\equiv (\delta \rho_i, \delta v_{x,i}, \delta v_{y,i}, \delta B_x, \delta B_y, \delta \rho_n, \delta v_{x,n},\delta v_{y,n})^T$  multiplied by $\exp[\rmi (k_xx+k_yy+\omega t)]$ under the WKBJ approximation, where $k_x$ is the longitudinal wavenumber, $k_y$ is the transverse wavenumber, and $\omega$ is the eigenvalue evaluated in the shock frame. By substituting these perturbations and the background states into the equations (\ref{eq:1})--(\ref{eq:5}), the following linearized equations are obtained (see \citet{Gu04} and \citet{GC20} in the 1D case):

\begin{equation}
P U = \rmi \omega U,\label{eq:disp_matrix_2D}
\end{equation}
where
\begin{eqnarray}
P=\left[\arraycolsep=6pt
\begin{array}{cccccccc}
-\rmi k_x V_i-2\beta \rho_i& -\rmi k_x \rho_i  & -\rmi k_y \rho_i & 0 & 0 & \xi_\mathrm{CR} & 0 & 0\\
-\rmi k_x \frac{c^2_s}{\rho_i} & -\rmi k_x V_i -\gamma \rho_n & 0 &  \rmi k_y \frac{V^2_{A,i}}{B_y} & -\rmi k_x \frac{V^2_{A,i}}{B_y}  & -\gamma V_d & \gamma \rho_n & 0\\
-\rmi k_y \frac{c^2_s}{\rho_i} & 0 & -\rmi k_x V_i -\gamma \rho_n & 0 & 0 & 0 & 0 & \gamma \rho_n \\
0 & \rmi k_y B_y & 0 & -\rmi k_x V_i & 0 & 0 & 0 & 0\\
0 & -\rmi k_x B _y & 0 & 0 & -\rmi k_x V_i & 0 & 0 & 0\\
0 & 0 & 0  & 0 & 0 & -\rmi k_x V_n  & -\rmi k_x \rho_n  & -\rmi k_y \rho_n\\
\gamma V_d & \gamma \rho_i & 0 & 0 & 0 & -\rmi k_x \frac{c_s^2}{\rho_n}  & -\rmi k_x V_n-\gamma \rho_i  & 0\\
0 & 0 & \gamma \rho_i & 0 & 0 & -\rmi k_y  \frac{c_s^2}{\rho_n}  & 0 & -\rmi k_x V_n -\gamma \rho_i
\end{array}
\right].\label{eq:C_ps}
\end{eqnarray}

\comment{
All the background gradients are ignored in the above linear equations due to the WKB approximation. However, some of the perturbed terms multiplied by the background gradients are unclearly insignificant. By retaining these terms, the above linear equations become
\begin{equation}
C U = \rmi \omega U,\label{eq:disp_matrix_2D_bg}
\end{equation}
where
\begin{eqnarray}
C=\left[\arraycolsep=6pt
\begin{array}{cccccccc}
-\rmi k_x V_i-2\beta \rho_i& -\rmi k_x \rho_i  & -\rmi k_y \rho_i & 0 & 0 & \xi_\mathrm{CR} & 0 & 0\\
-\rmi k_x \frac{c^2_s}{\rho_i} & -\rmi k_x V_i -\gamma \rho_n & 0 &  \rmi k_y \frac{V^2_{A,i}}{B_y} & -\rmi k_x \frac{V^2_{A,i}}{B_y}  & -\gamma V_d & \gamma \rho_n & 0\\
-\rmi k_y \frac{c^2_s}{\rho_i} & 0 & -\gamma \rho_n & {V_{A,i}^2 \over B_y}{d\ln B_y \over dx} & 0 & 0 & 0 & \gamma \rho_n \\
0 & \rmi k_y B_y & 0 & -\rmi k_x V_i & 0 & 0 & 0 & 0\\
0 & -\rmi k_x B _y & -V_i {d\ln B_y \over dx} & V_i   {d\ln B_y \over dx} & -\rmi k_x V_i & 0 & 0 & 0\\
0 & 0 & 0  & 0 & 0 & -\rmi k_x V_n  & -\rmi k_x \rho_n  & -\rmi k_y \rho_n\\
\gamma V_d & \gamma \rho_i & 0 & 0 & 0 & -\rmi k_x \frac{c_s^2}{\rho_n}  & -\rmi k_x V_n-\gamma \rho_i  & 0\\
0 & 0 & \gamma \rho_i & 0 & 0 & -\rmi k_y  \frac{c_s^2}{\rho_n}  & 0 & -\gamma \rho_i
\end{array}
\right].
\end{eqnarray}
}
Without the loss of generality, we adopt a constant wavenumber $k_x$ of $1/0.015$ pc$^{-1} (\equiv k_{fid}$), as used in \citet{GC20}. The choice of $k_x$ is made to satisfy the WKBJ approximation, as demonstrated in the right panel of Figure~\ref{fig:fig1} where $k_x L_p$ and $k_x L_B$ are shown to be all much larger than unity \citep[also refer to Figure 1 in][]{GC20}.

\begin{figure}
\plottwo{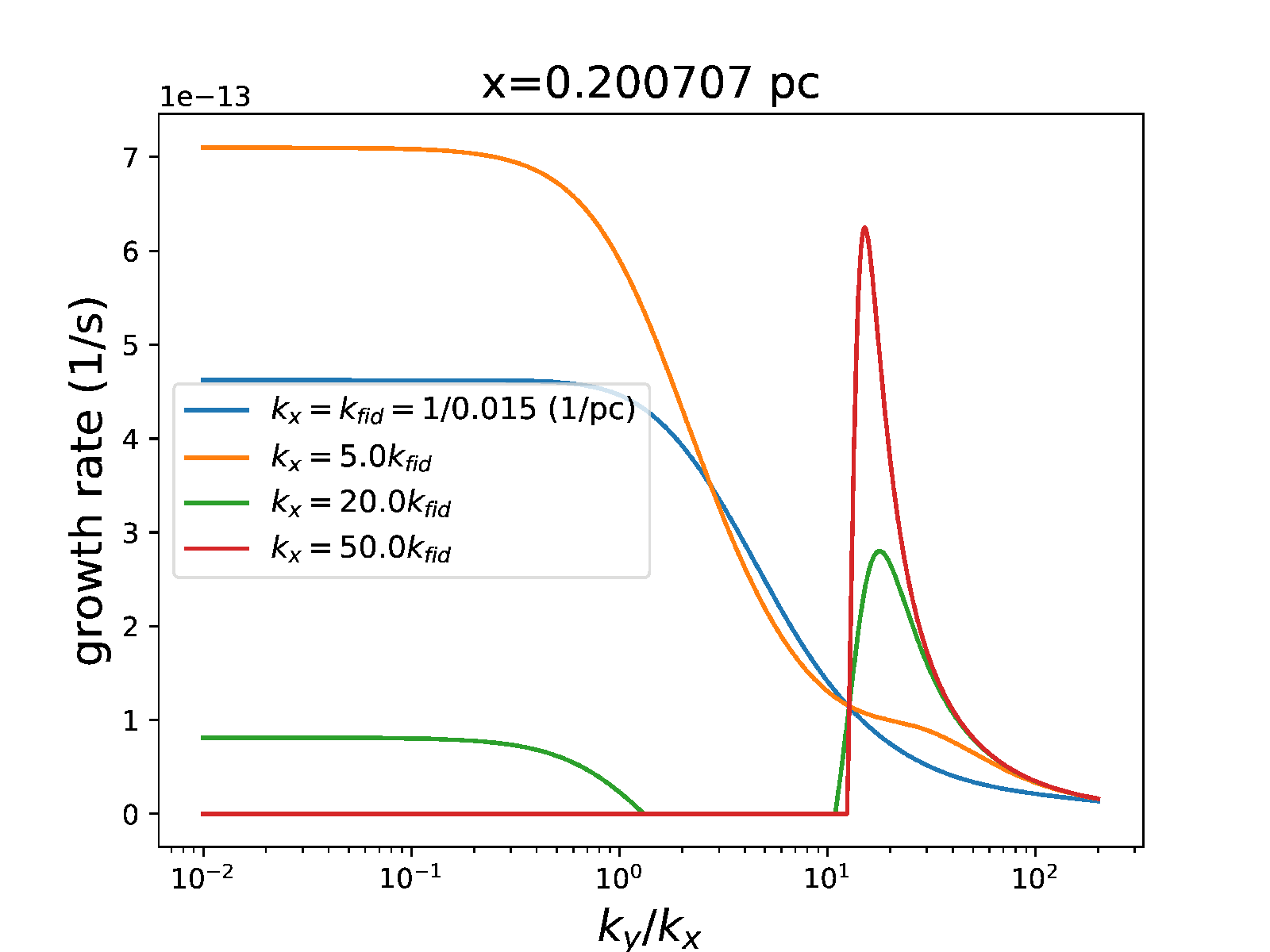}{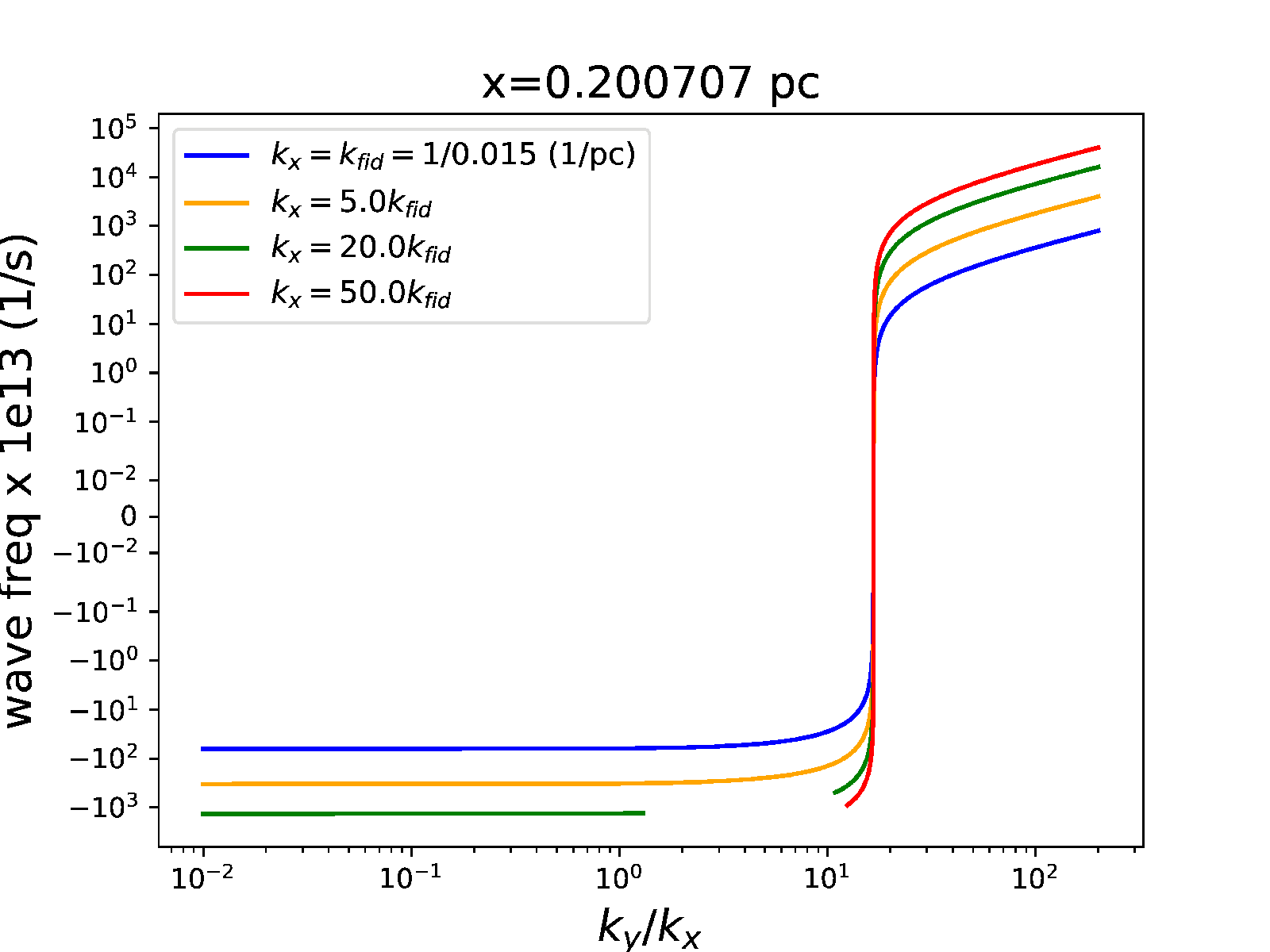}
\caption{Growth rate $\Gamma_{grow}$ (left panel) and its corresponding wave frequency $\omega_{wave}$ (right panel) of the drag instability as a function of $k_y/k_x$ in the four cases of $k_x$ at $x\approx 0.2$ pc. Note that the wave frequency undergoes a steep change from a negative to a positive value at $k_y/k_x \approx 16.84$ regardless of the value of $k_x$. The wave frequency is not presented when the growth rate is zero, i.e., the green curve in the range of $k_y/k_x \sim 1-10$ and the red curve in the range of $k_y/k_x \lesssim 10$.}
\label{fig:ps}
\end{figure}

Figure~\ref{fig:ps} shows the growth rate $\Gamma_{grow}$ ($=$Im[$\omega]<0$, left panel) and its corresponding wave frequency  $\omega_{wave}$ ($=$Re[$\omega]$, right panel) of the unstable wave mode as a function of $k_y/k_x$ in the shock frame, at the location of $x\approx 0.2$ pc in the middle of the C-shock. It is the only unstable mode among all eigenmodes computed from Equation(\ref{eq:disp_matrix_2D}). When the  longitudinal  wavelengths are not exceptionally short (i.e., $k_x=k_{fid}$ and $5k_{fid}$ in the figure), the growth rate decreases approximately with increasing $k_y/k_x$. However, when $k_x$ is as large as $20k_{fid}$ (green curve) and $50k_{fid}$ (red curve), the growth rate decreases and even drops to zero, except when $k_y/k_x$ is sufficiently large in this particular example, which corresponds to a jump of the wave frequency around $k_y/k_x \gtrsim 12.7$, as shown  in the right panel of Figure~\ref{fig:ps}. We explore the underlying physics for the mode properties with simplified dispersion relations in the following subsection.

\subsection{Simplified Dispersion Relations}
\label{sec:dispersion}

The eigenvalue problem for Equation(\ref{eq:disp_matrix_2D}) amounts to a complicated equation containing a polynomial of $\omega$ to the eighth power. To make the physical analysis of the unstable mode extractable, we attempt to numerically solve Equation(\ref{eq:disp_matrix_2D}) by removing as many terms as possible in the matrix $P$  but still obtain the same growth rate and wave frequency as those shown in Figure~\ref{fig:ps}. After this exercise, we realize that Equation(\ref{eq:disp_matrix_2D}) can be further reduced to
\begin{eqnarray}
(\Gamma_i + 2 \beta \rho_i){\delta \rho_i \over \rho_i} + \rmi k_y \delta v_{i,y} - \xi_{CR} {\rho_n \over \rho_i}{\delta \rho_n \over \rho_n} =0, \label{eq:redu1}\\
(\Gamma_i + \gamma \rho_n) \delta v_{i,y} - \gamma \rho_n \delta v_{n,y} =0,\label{eq:redu2}\\
\Gamma_n {\delta \rho_n \over \rho_n} + \rmi k_x \delta v_{n,x} + \rmi k_y \delta v_{n,y}=0, \label{eq:redu3}\\
-\gamma \rho_i V_d {\delta \rho_i \over \rho_i} + \rmi k_x c_s^2 {\delta \rho_n \over \rho_n} + (\Gamma_n + \gamma \rho_i) \delta v_{n,x} =0,\label{eq:redu4}\\
-\gamma \rho_i \delta v_{i,y} + \rmi k_y c_s^2 {\delta \rho_n \over \rho_n} + (\Gamma_n + \gamma \rho_i) \delta v_{n,y}=0, \label{eq:redu5}
\end{eqnarray}
where $\Gamma_i = \rmi (\omega + k_x V_i)$ is the eigenvalue in the comoving frame of  the ions and $\Gamma_n=\rmi (\omega + k_x V_n)$ is the eigenvalue in the comoving frame of the neutrals.
Since the collisional rate for an ion with the ambient neutrals $\gamma \rho_n$ is tremendously large in a weakly ionized cloud, Equation(\ref{eq:redu2}) implies that $\delta v_{i,y} \sim \delta v_{n,y}$, which allows Equation(\ref{eq:redu5}) to be further simplified to 
\begin{equation}
\rmi k_y c_s^2 {\delta \rho_n  \over \rho_n} + \Gamma_n \delta v_{n,y} \approx 0, \label{eq:redu5_1}
\end{equation}
and consequently, Equation~(\ref{eq:redu1}) can be approximated to
\begin{equation}
(\Gamma_i + 2 \beta \rho_i){\delta \rho_i \over \rho_i} +  {k_y^2 c_s^2 \over \Gamma_n} {\delta \rho_n \over \rho_n} - \xi_{CR} {\rho_n \over \rho_i}{\delta \rho_n \over \rho_n} \approx 0. \label{eq:redu1_1}
\end{equation}
The absence of the induction equations for $\delta {\bf B}$ and the momentum equation for $\delta v_{i,x}$ in the above reduced set of linear equations (\ref{eq:redu1})--(\ref{eq:redu5}) indicates that the disturbance of the ion flow along the shock direction and that of magnetic fields plays a minor role in the dynamical evolution of the 2D drag instability.

In a 2D perpendicular shock, the linearized equations should approach the 1D result when $k_y \ll k_x$ (i.e., a transversely large-scale mode). 
The drag instability can occur in a 1D steady C-shock \citep{GC20}. 
Specifically, for the definite occurrence of the instability in 1D,
the rate of the mode $\Gamma_n$ observed in the comoving frame of the neutrals is considerably smaller than both the recombination rate $2 \beta \rho_i$ and the ion-neutral drift rate across a distance of one wavelength (i.e., $k|V_d|$), whereas it is considerably larger than the neutral collision rate with the ions (i.e., $\gamma \rho_i$) and the sound-crossing rate over one wavelength (i.e., $k_x c_s$).
When the aforementioned conditions are satisfied, the 2D linearized equations (i.e., Equations(\ref{eq:redu3}), (\ref{eq:redu4}), \& (\ref{eq:redu1_1})) are reduced to those for a 1D perpendicular shock, which yield the simplified dispersion relation  \citep{Gu04,GC20}
\begin{equation}
\Gamma_n \approx \pm {(1+\rmi)\over 2} \sqrt{{k_xV_d} \gamma \rho_i}=\pm  {(1+\rmi)\over 2}  \sqrt{k_x \over L_B} V_{A,n}, \label{eq:disp_Gu}
\end{equation}
where the positive sign corresponds to a growing wave. As pointed out by \citet{GC20}, the wave frequency in the shock frame $\omega_{wave}$ is dominated by $k_x V_n$ instead of the imaginary part of $\Gamma_n$ due to the fast background flow across the shock width downstream.

On the other hand, in the regime where $k_y \gg k_x$ (i.e., a transversely small-scale mode) such that $k_y c_s$  is much larger than $2\beta \rho_i$, $k_x V_i$, and $k_x V_n$, Equation(\ref{eq:redu1_1}) suggests that the terms associated with ionization, recombination, and the background shock flows are less significant, revealing a mode with the wave frequency Re[$\omega$]$\approx k_y c_s$ much larger than Im[$\omega]$. Therefore,  Equation(\ref{eq:redu1_1}) suggests the relation $\delta \rho_i /\rho_i \approx \delta \rho_n/\rho_n$. In this regime, the in-phase relation between $\delta \rho_i$ and $\delta \rho_n$ is no longer maintained by the ionization--recombination equilibrium in the regime of a small $k_y/k_x$ but is a consequence of the fast acoustic wave along the background magnetic fields,\footnote{In comparison, $\delta \rho_i/\rho_i=(1/2)\delta \rho_n/\rho_n$ when the ionization--recombination equilibrium is attained. The factor $1/2$ does not appear in this regime when the slow mode is faster than the recombination process.} i.e., the so-called slow mode, because $c_s < V_{A,n}$ in our fiducial model \citep[e.g.,][]{Shu}. This result, along with Equations(\ref{eq:redu3}), (\ref{eq:redu4}), and (\ref{eq:redu5_1}), yield the following simplified dispersion relation:
\begin{equation}
\Gamma_n \approx \pm \left( -\rmi k_x |V_d| \gamma \rho_i -k^2 c_s^2 \right)^{1/2} \approx 
 \pm \left( {1\over 2} {k_x |V_d| \gamma \rho_i \over k_y^2 c_s^2} + \rmi \right) k_y c_s,\label{eq:disp_per}
\end{equation}
where $k^2 \equiv k_x^2+k_y^2$ and moreover, we have used $\Gamma_n^2 \gamma \rho_i \approx k_y^2 c_s^2 \gamma \rho_i$ from Equation(\ref{eq:redu1_1})  and $k_y c_s \gg \sqrt{k_x |V_d| \gamma \rho_i}$ in deriving the above equation.  The inequality $k_y c_s \gg \sqrt{k_x |V_d| \gamma \rho_i}$ holds in the regime where $k_y c_s \gg 2\beta \rho_i$ is being considered here, along with $2\beta \rho_i \gg \sqrt{k_x |V_d| \gamma \rho_i}$ in the fiducial model \citep[see the left panel of Figure 3 in][]{GC20}.\footnote{Recall that in a 1D C-shock, $\sqrt{k_x |V_d| \gamma \rho_i}$ is about the growth rate  of the drag instability (see Equation~\ref{eq:disp_Gu}), which is smaller than $2\beta \rho_i$ such that the ionization--recombination equilibrium can be attained for the perturbations to allow the instability to occur.}

The positive sign of Equation(\ref{eq:disp_per})
corresponds to a growing wave with the growth rate related to the ion-neutral drag and the wave frequency $k_y c_s$ associated with the slow mode. More specifically, in the comoving frame of the neutrals, the phase velocity of the unstable wave is about $-k_y c_s/k_x \hat x - c_s \hat y$ and the group velocity is approximately $-(k_x/k_y)c_s \hat x - c_s \hat y$. Hence, in this regime, while both the phase and signal of the unstable wave propagate mainly along the background field lines at the sound speed, the signal also propagates upstream slowly at a speed much smaller than the sound speed.
In the shock frame,  the phase velocity of the unstable wave is about $[-k_y c_s/k_x +V_n] \hat x +[- c_s + (k_x/k_y)V_n] \hat y \approx -k_y c_s/k_x \hat x - c_s \hat y$.

Based on the above simplified dispersion relations for unstable modes on different transverse scales,
we may be able to interpret the results 
%for the cases with moderate values of $k_x$ (i.e., $k_x=k_{fid}$ and 5$k_{fid}$) 
illustrated in Figure~\ref{fig:ps}. When $k_y/k_x \rightarrow 0$, both the growth rate and wave frequency converge to the 1D result for the drag instability \citep{GC20}, as illustrated by the flat curves for $k_y/k_x <1$ in Figure~\ref{fig:ps}. However, as $k_y/k_x$ increases, a new property of the growing mode associated with 2D emerges when the acoustic wave along the background field line, i.e. the slow mode, becomes important. This happens when $c_s k_y \sim k_xV_n$ as suggested by the transition of the dispersion relation from Equation(\ref{eq:disp_Gu}) to Equation(\ref{eq:disp_per}). Physically, it occurs when the slow-mode rate $c_s k_y$ introduced by the additional dimension along the background field in the $y$-direction becomes comparable to the shock-crossing rate $k_x V_n$ in the $x$-axis through the C-shock. Owing to this transition at the large $k_y$, the jump of the wave frequency from a negative to a positive value occurs at $k_y/k_x \sim V_n/c_s \equiv \hat k_{jump} \approx 16. 84$ at $x\approx 0.2$ pc, which agrees with the right panel of   Figure~\ref{fig:ps}.  While we show the value of $\hat k_{jump}$ at one particular $x$, the range of $\hat k_{jump}$ within a C-shock can be estimated by the pre-shock conditions. The $V_n$ decreases from $v_0$ at the beginning of a shock to $v_0/r_f$ at the end of a shock, where $r_f$ is the final compression ratio of the neutral density in the post-shock region given approximately by $\sqrt{2} v_0/v_{A,n,0}$ \citep{CO12}. Hence, $\hat k_{jump}$ decreases from $v_0/c_s$ to $(v_0/r_f)/c_s \approx B_0/(\sqrt{4 \pi \rho_{n,0}}c_s)$. In our fiducial model,
$\hat k_{jump}$ changes from about 25 to 3 across the shock width. In the typical environments of star-forming clouds, $v_0 \sim$ 1-6 km/s, $n_0 \sim$ 100-1000 1/cm$^3$, $B_0 \sim 5-10$ $\mu$G, and the temperature is about 10 K \citep[e.g., see Table 1 in][]{CO12}. Therefore, $\hat k_{jump}$ can decrease from $\sim$ 10-30 at the beginning of a C-shock to $\sim$ 1.5-5 at the end of a C-shock.

\begin{figure}
\plottwo{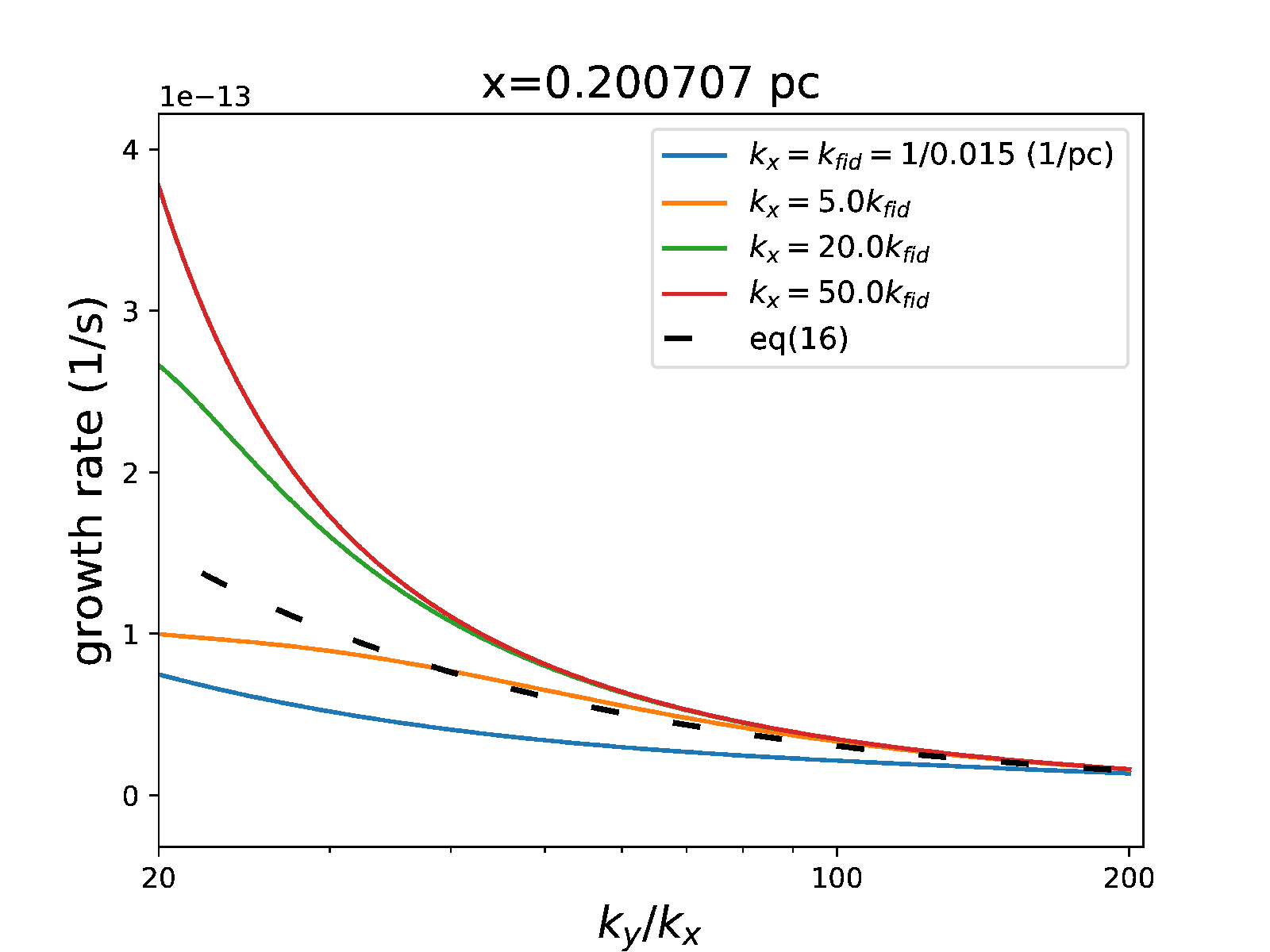}{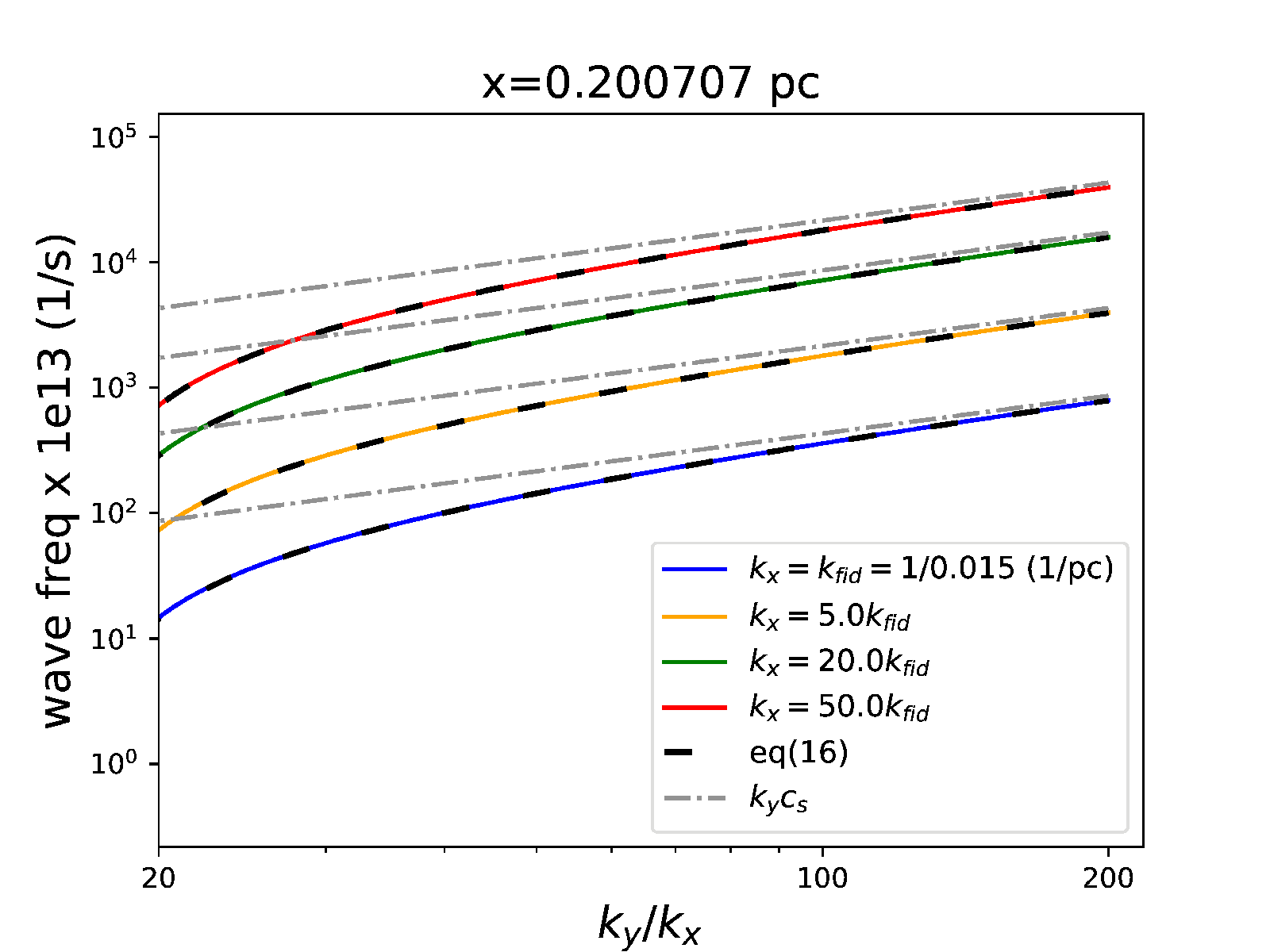}
\caption{Blowup of the region where $k_y/k_x >20$ in Figure~\ref{fig:ps} to compare the exact behavior of the growing mode with the approximate result from Equation(\ref{eq:disp_per}). The curves of $k_y c_s$ corresponding to various $k_x$ are also plotted on the right panel for comparison.}
\label{fig:ps_disper}
\end{figure}

As $k_y/k_x$ further increases to an even larger value, the behavior of the growing mode follows the dispersion relation described by Equation(\ref{eq:disp_per}). This is demonstrated in Figure~\ref{fig:ps_disper}, which zooms in to the region where $k_y/k_x > 20$ in Figure~\ref{fig:ps}. 
The dashed lines present the growth rate Re[$\Gamma_n$] (left panel) and the wave frequency  Im[$\Gamma_n-k_xV_n$] (right panel) calculated from Equation(\ref{eq:disp_per}). As $k_y/k_x$ increases, the exact growth rates of the unstable mode for various $k_x$ computed from Equation(\ref{eq:disp_matrix_2D})  converge to  Re[$\Gamma_n$] while the exact wave frequencies $\omega_{wave}$ overlap with  Im[$\Gamma_n-kV_n$] in the entire zoomed-in region. The slow-mode frequency  $k_y c_s$, as illustrated by the dashed-dotted lines, is also plotted for each value of $k_x$. As $k_y/k_x$ increases, the wave frequency $\omega_{wave}$ in all cases converges toward $k_y c_s$, confirming that the slow mode emerges and predominates on small transverse scales for the drag instability.
Note that the positive wave frequency for $k_y/k_x > \hat k_{jump}$ indicates a growing wave propagating upstream within the 2D perpendicular C-shock, in stark contrast to the growing mode for $k_y/k_x < \hat k_{jump}$, which is more dynamically dominated by the wave convected with the downstream flow across the shock width, similar to the result for a 1D C-shock.

In the case of a moderate value of $k_x$ (i.e., $k_x=k_{fid}$ and $5k_{fid}$), the growth rate decreases  almost monotonically with increasing $k_y$ when $k_y/k_x \gtrsim \hat k_{jump}$, in an approximate agreement with the trend described by Equation(\ref{eq:disp_per}). On the other hand, 
 in the regime where $k_y/k_x <1$,  the growth rate is small or nearly zero when $k_x$ is large enough (i.e., $k_x=50$ and $100 k_{fid}$ in Figure~\ref{fig:ps}) for the small-scale gas pressure perturbation along the shock direction to suppress the growth \citep{Gu04,GC20}. However, a growing mode exists when $k_y/k_x \gtrsim 12.7$. Its growth rate rises significantly during the wave frequency transition and then declines, according to Equation(\ref{eq:disp_per}), as $k_y/k_x$ increases to a large value. We are unable to derive a simple dispersion relation to describe the rise of the growth rate during the wave frequency transition. It is because more terms in Equations(\ref{eq:redu1})--(\ref{eq:redu5})
 become comparably important and thus cannot be neglected as a proper approximation for deriving Equation(\ref{eq:disp_per}). 
 Nevertheless, this complexity of the distinct behaviors of the growth rate at the wave frequency transition can be unraveled by studying the phase differences between perturbed quantities, which will be presented in the next subsection.
 
 \subsection{An auxiliary analysis based on phase differences between perturbations}

\begin{figure}
    \centering
    \subfigure[]{\includegraphics[width=0.32\textwidth]{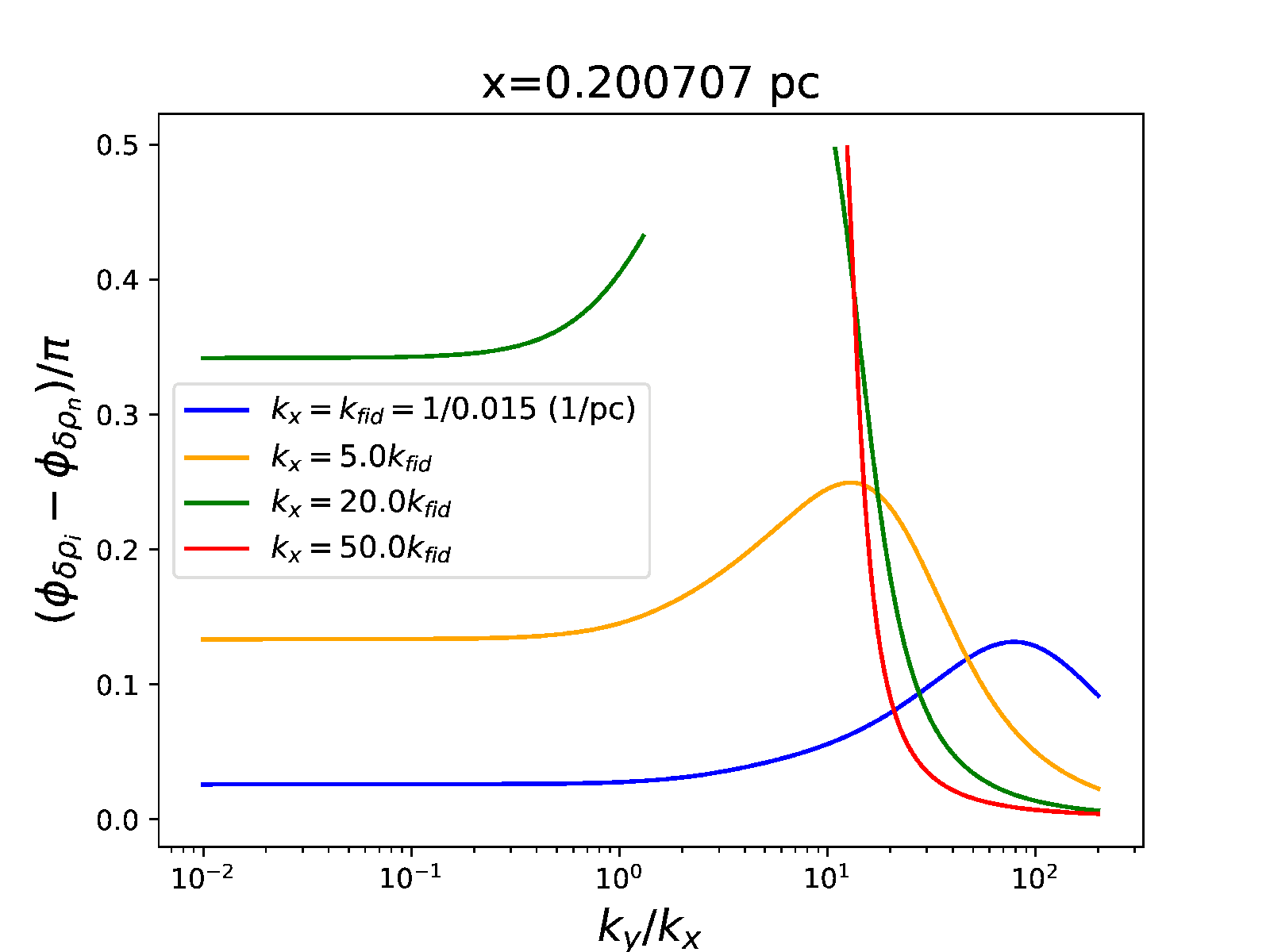}} 
    \subfigure[]{\includegraphics[width=0.32\textwidth]{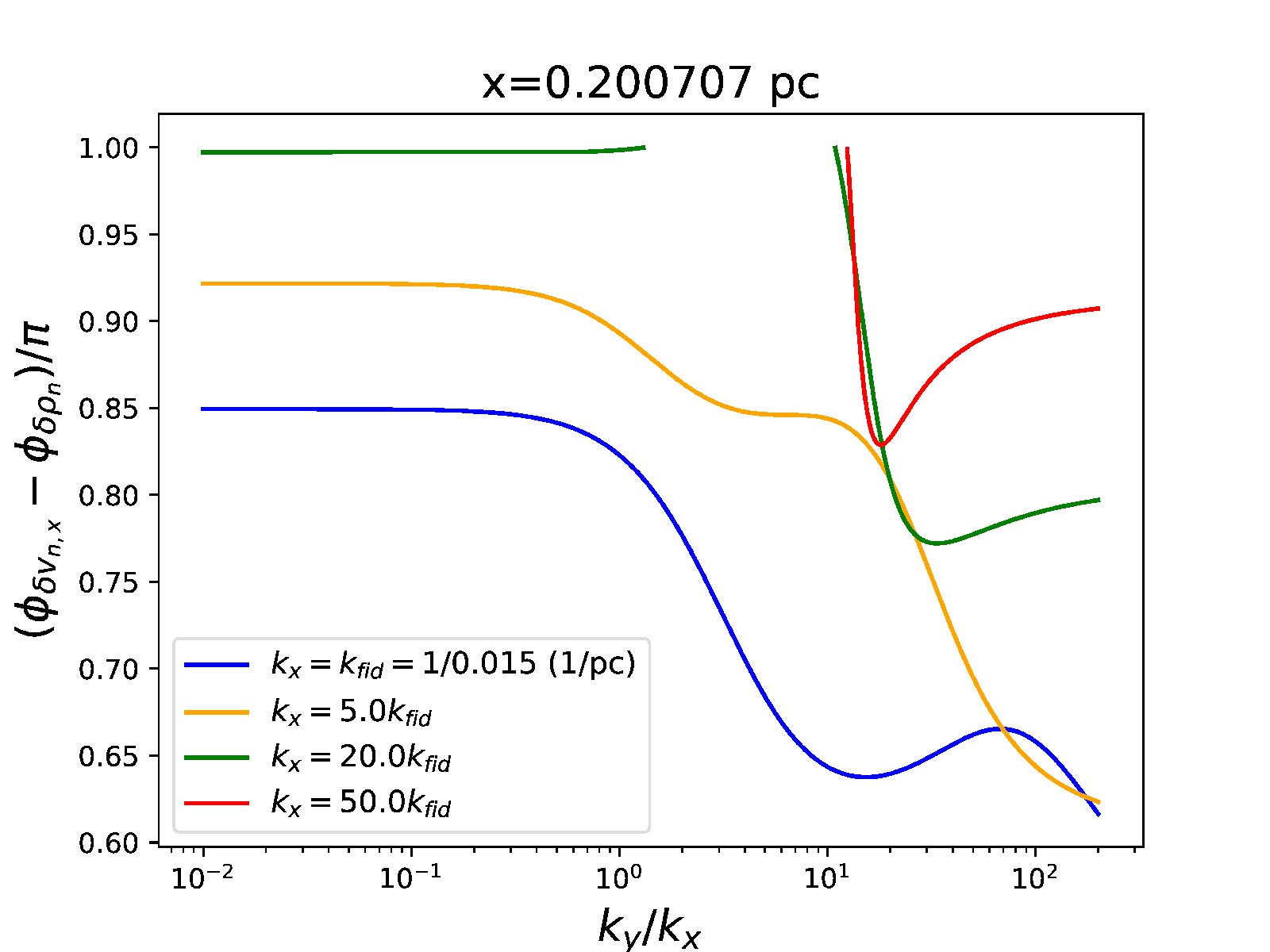}} 
    \subfigure[]{\includegraphics[width=0.32\textwidth]{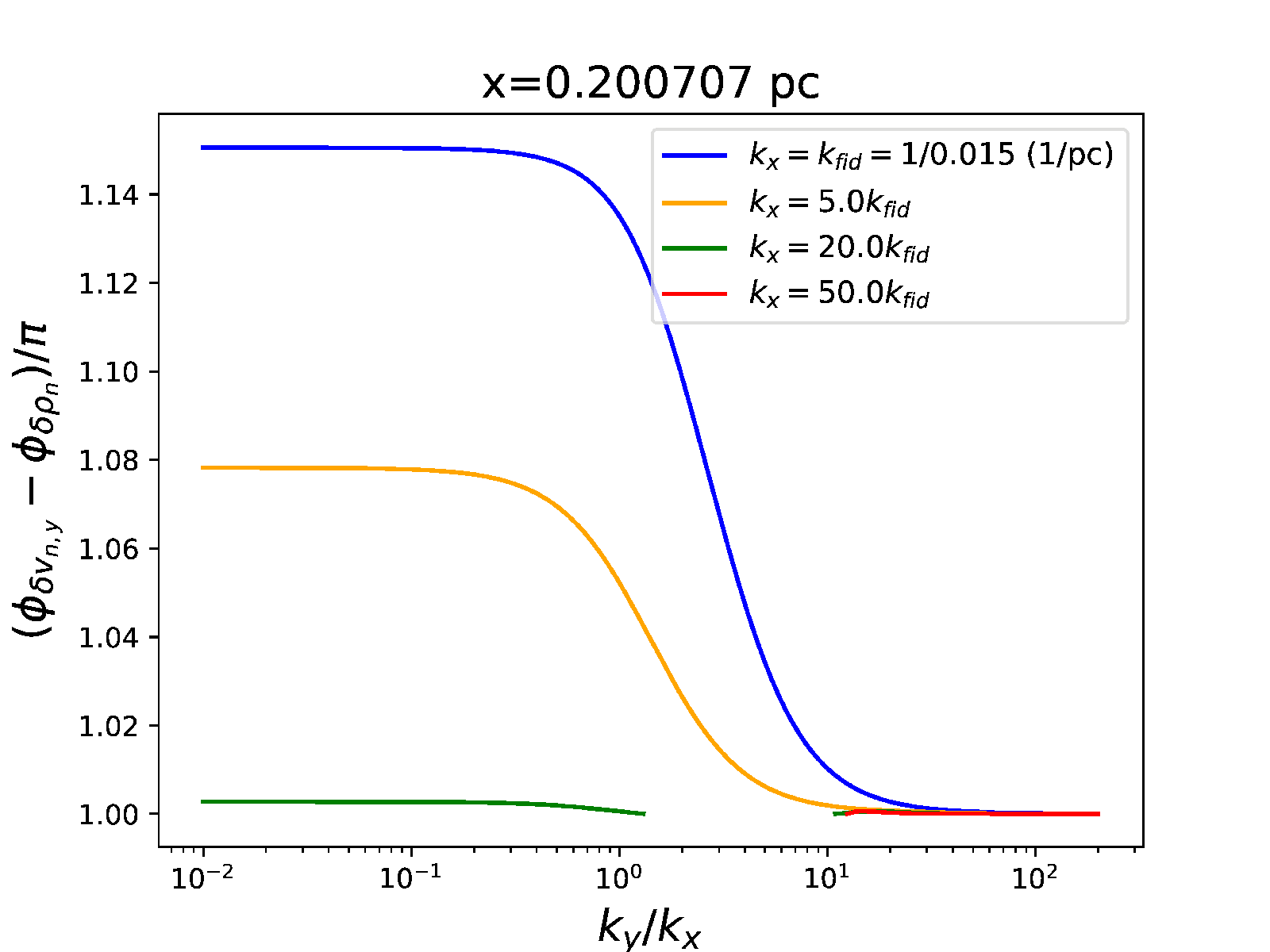}}
    \caption{Phase differences of unstable modes between $\delta \rho_i$ and $\delta \rho_n$ (panel (a)),  $\delta v_{n,x}$ and $\delta \rho_n$ (panel (b)), and $\delta v_{n,y}$ and $\delta \rho_n$ (panel (c)) as a function of $k_y/k_x$ at the location of $x\approx 0.2$ pc in the fiducial model of the perpendicular C-shock. Note that there is a steep change in the green and red curves during the wavelength transition at $k_y/k_x \approx  \hat k_{jump} = 16.84$ in panels (a) and (b). The phase differences are not presented when the growth rates are zero for some values of $k_y/k_x$ (refer to the left panel of Figure~\ref{fig:ps}).}
    \label{fig:ph_diff}
\end{figure}

We illustrate the phase differences in Figure~\ref{fig:ph_diff} corresponding to the growth rate and wave frequency shown in Figure~\ref{fig:ps}.  Figure~\ref{fig:ph_diff} includes the phase difference
between the density perturbations, as well as those between the neutral velocity and density perturbations. The flat part of the curves in the range of $k_y/k_x \lesssim 1$  corresponds to the flat part of the curves in Figure~\ref{fig:ps}, resembling the 1D drag instability for the 2D unstable modes. 
In contrast to the cases for the small $k_x$ (i.e., blue and orange curves), the phase difference between $\delta \rho_i$ and $\delta \rho_n$ in the cases for the large $k_x$ (green and red curves in panel (a) of Figure~\ref{fig:ph_diff}) exhibit a dramatic change when $k_y/k_x \approx \hat k_{jump}=16.8$, corresponding to the jump of the wave frequency. Since the wave frequency of the mode $\omega_{wave}$ is small around this jump transition, the Doppler-shift frequency $k_x V_i$ with a large $k_x$ becomes nonnegligible compared to the ionization rate in the continuity equation of the ions.
 Consequently,  the phase difference between the ion and neutral density perturbations does not stay small by the ionization--recombination equilibrium but
 becomes large.  The phase difference decreases steeply with increasing $k_y/k_x$ because the rate of the mode is predominated quickly by the slow mode $\approx k_y c_s$, resulting in the small phase difference between the ion and neutral density perturbations for a large $k_y/k_x$. It is in accordance with the result that $\delta \rho_i/\rho_i \approx \delta \rho_n/\rho_n$ when deriving  Equation(\ref{eq:disp_per}). 
 
 The drastic change in $\phi_{\delta \rho_i}-\phi_{\delta \rho_n}$ also causes the steep change in $\phi_{\delta v_{n,x}}-\phi_{\delta \rho_n}$ from $\pi$ to $\approx 0.8\pi$ around the frequency transition (see the green and red curves in panel (b) of Figure~\ref{fig:ph_diff}) through the linearized continuity and momentum equations, leading to a bump in the growth rate of the drag instability near the frequency transition, as shown in the left panel of Figure~\ref{fig:ps}.\footnote{Recall that the drag instability in a 1D case is driven by 
 the canonical phase difference $3\pi/4$ between $\phi_{\delta v_{n,x}}$ and $\phi_{\delta \rho_n}$  \citep{Gu04,GC20}.}
 Since an acoustic mode can be characterized by the phase difference $\pi$ between density and velocity perturbations, the nearly out-of-phase difference between $\delta v_{n,y}$ and $\delta \rho_n$ shown in panel (c) of Figure~\ref{fig:ph_diff} for $k_y/k_x \gtrsim16.8$ reconfirms the right panel of Figure~\ref{fig:ps_disper}, i.e., the emergence of the slow mode when $k_y/k_x \gtrsim \hat k_{jump}$. It also agrees with Equation(\ref{eq:redu5_1}) on the relation $\Gamma_n \sim \rmi k_y c_s$ for $k_y/k_x > \hat k_{jump}$, which leads to $\delta \rho_n/\rho_n \sim - \delta v_{n,y}/c_s$, i.e., out of phase between $\delta \rho_n$ and $\delta v_{n,y}$ for a slow mode.

\subsection{Substantial growth of slowly propagating waves}
\label{sec:fast}

It was shown by \citet{GC20} that the growth of the drag instability in a 1D perpendicular shock is limited by the short time span for an unstable wave to stay within a C-shock as the wave is convected downstream by a fast flow across the shock; specifically, the local growth rate $\Gamma_{grow}$ is much smaller than the wave frequency in the shock frame $\omega_{wave}$. Consequently, a stronger shock with a larger shock width favors a more appreciable growth of the 1D drag instability in the typical environments of star-forming clouds investigated by the authors. However, for a 2D perpendicular shock, 
as has been studied in the preceding subsection, the transition of the wave frequency from a negative to a positive value allows for
the presence of a slowly traveling mode with an arbitrarily small wave frequency in the shock frame when $k_y/k_x \sim \hat k_{jump}$, thereby allowing a tremendous amount of time for the growth of the drag instability within a shock width in this transition regime. 

\begin{figure}
\plottwo{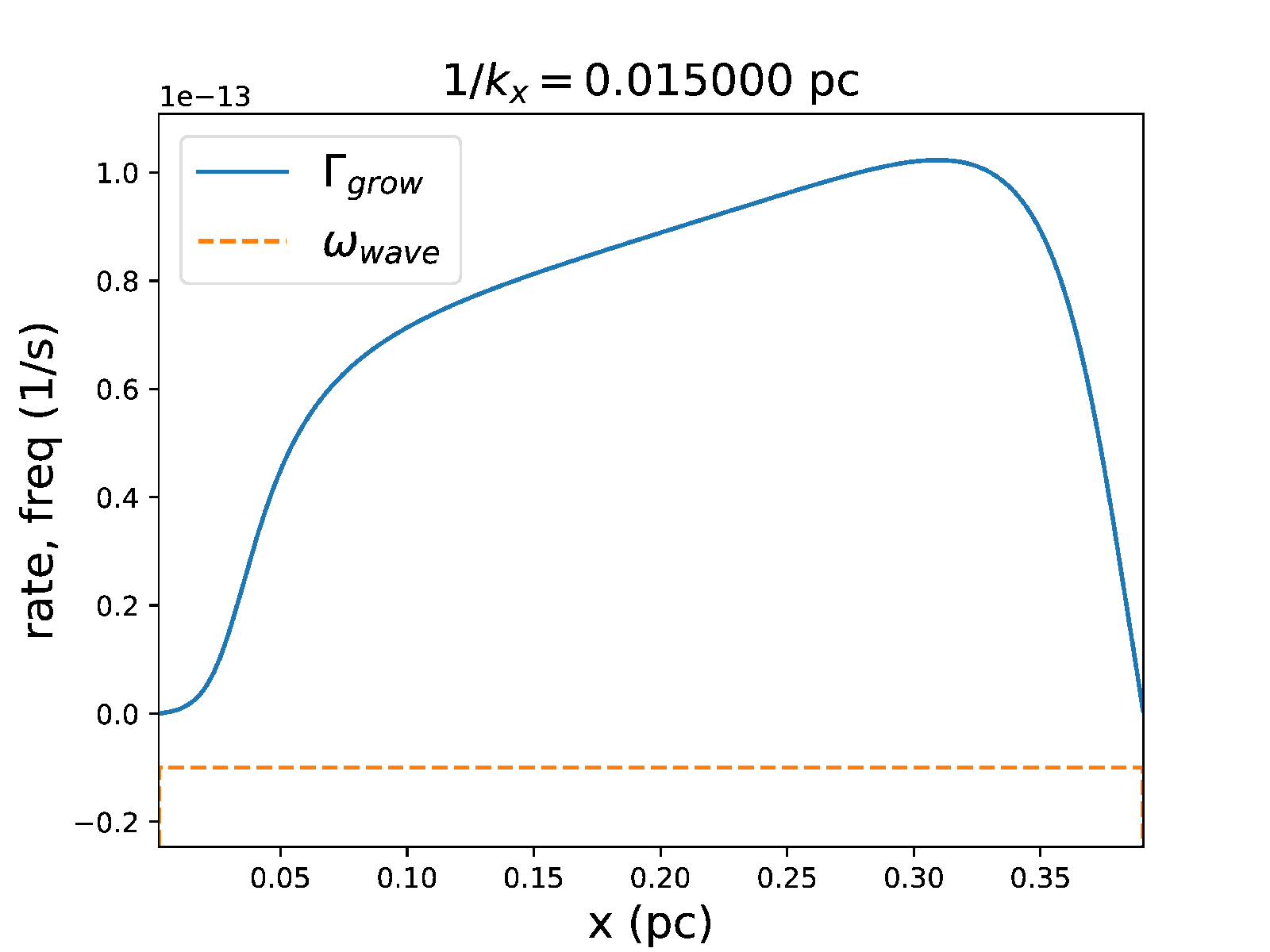}{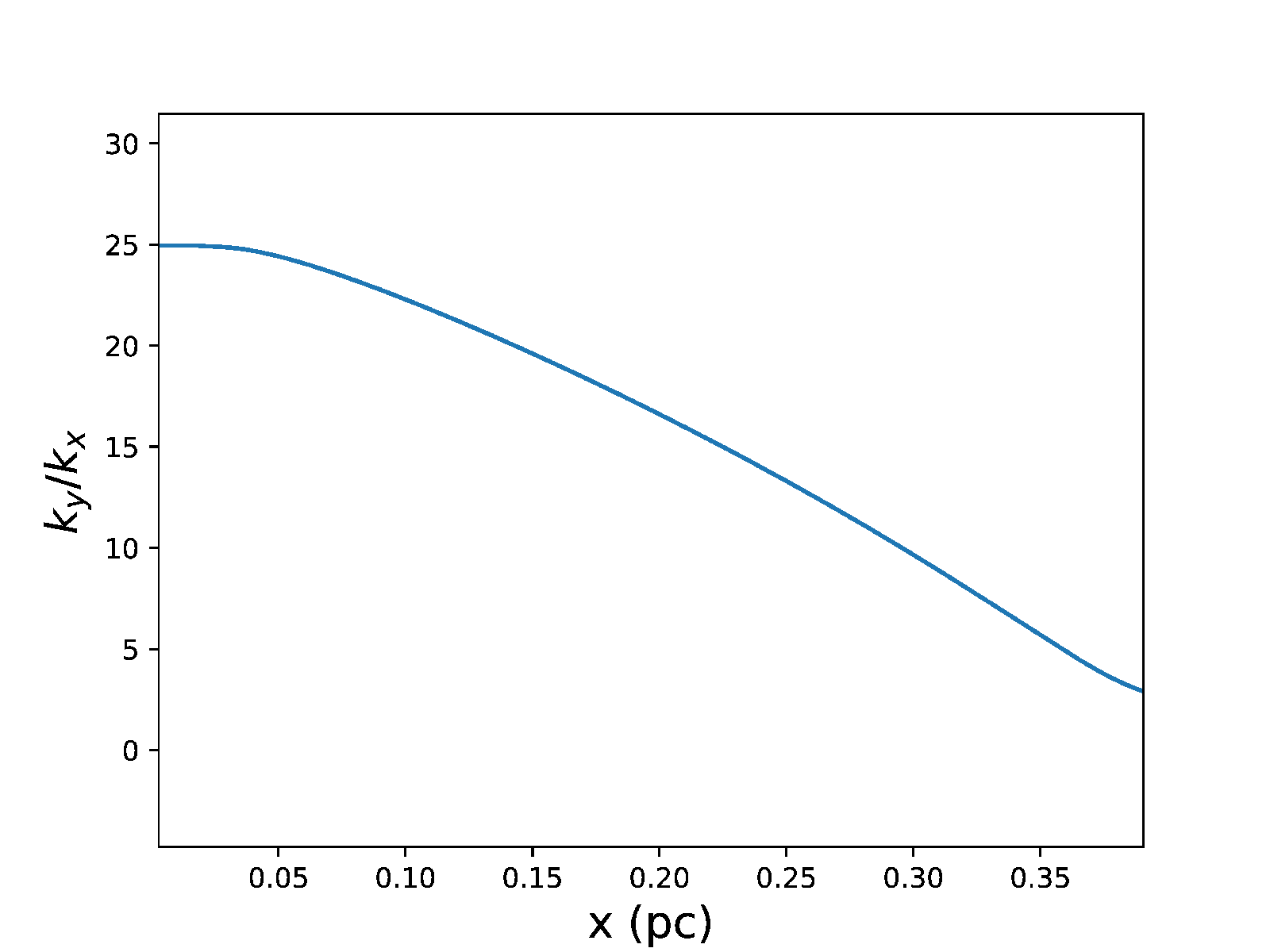}
\caption{Growth rate $\Gamma_{grow}$ as a function of $x$ for the unstable mode with the wave frequency $\omega_{wave}=-10^{-14}$ 1/s and $1/k_x=0.015$ pc (left panel) and the profile of the corresponding wavenumber ratio $k_y/k_x$ (right panel) for the fiducial model of the perpendicular C-shock.}
\label{fig:mode}
\end{figure}

We investigate this expectation by simply considering a mode with a small wave frequency $\omega_{wave}=-10^{-14}$ 1/s $\ll$ the flow crossing time over one  longitudinal  wavelength $k_x V_n$. According to the eigenvalue problem, a family of unstable modes should exist with the proper combination of the wavenumbers $k_x$ and $k_y$ as a function of $x$. For the purpose of computational convenience without loss of generality, we first consider the uniform $k_x$ given by the fiducial wavenumber $k_x=k_{fid}=$ 1/(0.015 pc), while allowing the $k_y$ of the mode to vary with $x$. The results are shown in Figure~\ref{fig:mode}. The resulting growth rate $\Gamma_{grow}$  is indeed larger than $\omega_{wave}$ in most of the region within the C-shock, as illustrated in the left panel of Figure~\ref{fig:mode}. The corresponding profile of $k_y/k_x$ decreases from 25 to 3 with increasing $x$, in agreement with the transition wavenumber ratio $k_y/k_x \approx \hat k_{jump}=V_n/c_s$, where $V_n$ decreases across the shock width due to density compression, as has been explained in Section~\ref{sec:dispersion}.
%as shown in Figure~\ref{fig:ps}. In fact, $k_y/k_x\approx 16.77$ is close to $V_n/c_s \approx 16.84$ at $x=0.2$ pc in Figure~\ref{fig:mode}. 
Next, we consider the uniform $k_y$ given by $15k_{fid}$ (i.e. a transversely small-scale mode) while allowing the $k_x$ of the mode to vary with $x$. The results of the growth rate and wave frequency are considerably similar to those shown in Figure~\ref{fig:mode}, affirming the slow propagation of the unstable modes with $k_y/k_x \approx \hat k_{jump}$ in a 2D perpendicular C-shock. We also study the ``counterpart" mode with the small positive wave frequency $\omega_{wave}=10^{-14}$ 1/s for the same case of either the uniform $k_x$ or uniform $k_y$. We find that the profiles of the growth rate and $k_y/k_x$ are almost identical to those presented in Figure~\ref{fig:mode}, except that the unstable wave slowly propagates upstream rather than downstream due to the sign change of $\omega_{wave}$.

As the unstable wave propagates slowly at the phase velocity $v_{ph,x}$, the unstable mode is nearly comoving with the shock during the cloud lifetime, which is typically about  tens of millions of years \citep{Engargiola,Blitz,Kawamura,Murray,Miura,Meidt,JK18}. Consequently, the MTG is no longer a proper measure for the growth of the slowly propagating modes in the shock frame. For instance, it takes about 82 million years for the unstable mode that we have considered with $\omega_{wave}=-10^{-14}$ 1/s and uniform $k_x=k_{fid}$ to travel across the entire shock width,\footnote{The wave-crossing time through the shock width is given by $\int_{\rm shock\  width} dx /|v_{ph,x}| = \int dx / (|\omega_{wave}|/k_x)$.} which is $\gtrsim$ the cloud lifetime. It is evident from Figure~\ref{fig:ps} that there are modes that are almost/exactly stationary in the shock frame (i.e., $\omega_{wave} \approx 0$) and have $\Gamma_{grow} \sim 10^{-13}$ 1/s $\sim 1/0.3$ 1/Myr. Hence, the growth of these modes is approximately given by $\exp(\Gamma_{grow} t)$, which can be substantially larger than unity if  $t$ is some fraction of the cloud lifetime. In contrast, the maximum growth of the unstable mode in the 1D case is 
limited by the shock width and thus is given by MTG, which is merely about 9.9  in the fiducial model \citep{GC20}. The 2D mode ought to grow to a nonlinear phase according to the expectation from the linear theory.

\begin{figure}
\plotone{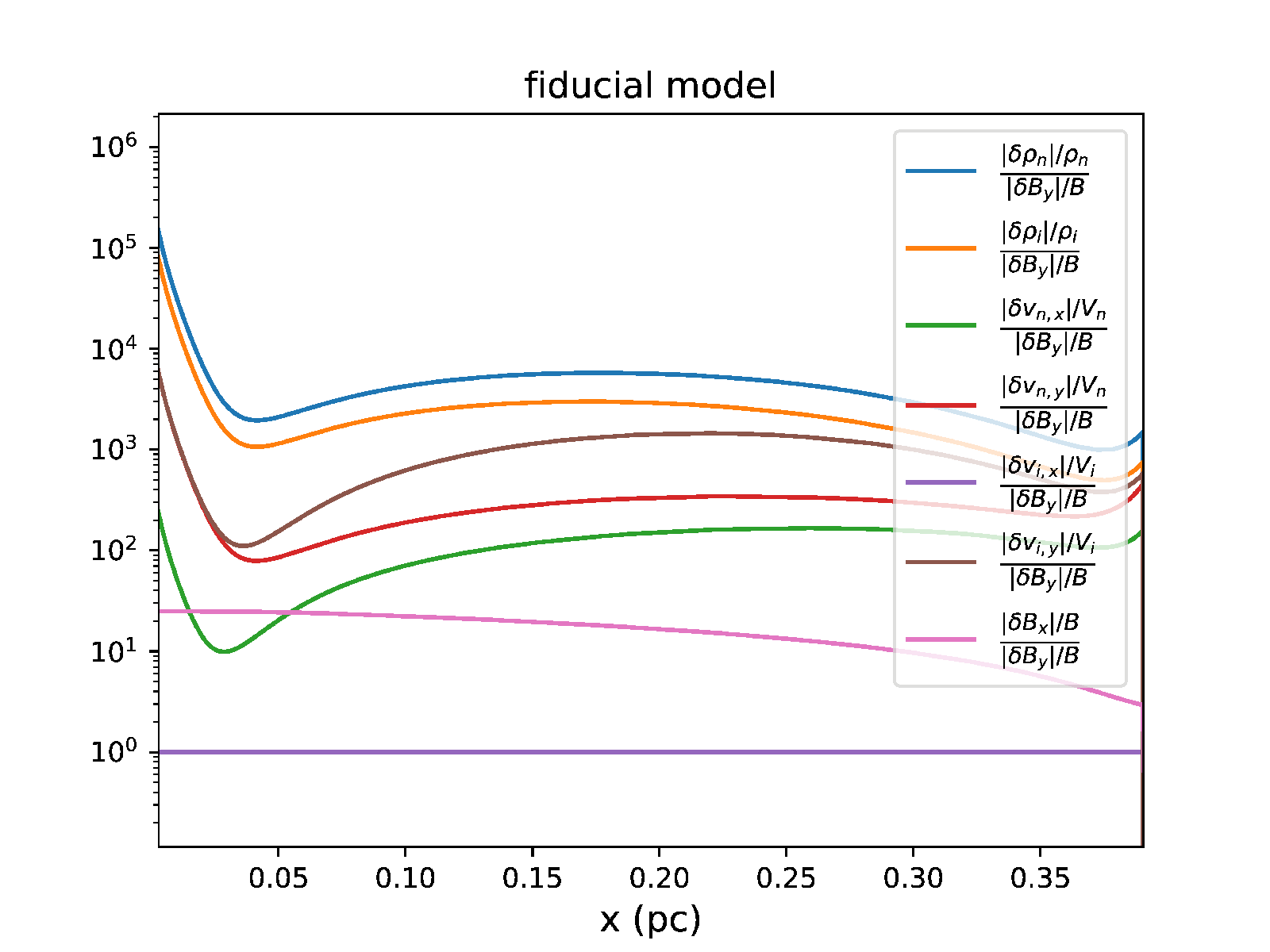}
\caption{Amplitude of perturbations normalized by the $y$-component of the magnetic perturbation $|\delta B_y|/B$ for the unstable mode presented in Figure~\ref{fig:mode}.}
\label{fig:eigenv}
\end{figure}

Figure~\ref{fig:eigenv} shows the amplitude of perturbations for the unstable mode presented in Figure~\ref{fig:mode}. The perturbed quantities are normalized by the $y$-component of the magnetic field perturbation $|\delta B_y|/B$. It is evident from the figure that within the shock width where the unstable mode exists, the density perturbations $\delta \rho_n/\rho_n$ and $\delta \rho_i/\rho_i$ dominate over the others, implying that the growth of density perturbations plays a decisive role in the dynamical evolution of the instability. Although this outcome of promoting the local density growth of a wave turns out to be the same as that in the 1D case \citep{GC20}, the drag instability is coupled with the slow mode in the regime $k_y/k_x > \hat k_{jump}$ for a 2D perpendicular shock, as described in the previous section. This can also be realized from Figure~\ref{fig:eigenv}, which shows $|\delta v_{n,y}| \gg |\delta v_{n,x}|$ and $|\delta v_{i,y}| \gg |\delta v_{i,x}|$ as a result of the presence of the slow mode propagating along the background magnetic fields in the $y$-direction. 

It is worth noting that $\delta B_x$ and $\delta B_y$ are generated from $B$ due to the gradients of the  longitudinal  motion $\partial_y (\delta v_{i,x})$  (i.e., magnetic wiggling) and  $\partial_x (\delta v_{i,x})$ (i.e., magnetic compression), respectively, through the induction equation such that $|\delta B_x|/|\delta B_y|=k_y/k_x$, in agreement with the ratios  $|\delta B_x|/|\delta B_y|$ in Figure~\ref{fig:eigenv}  and $k_y/k_x$ in the right panel of Figure~\ref{fig:mode}. However, given their small amplitudes, magnetic field perturbations simply grow passively through the induction equation and do not play any significant role in the dynamics of the drag instability, confirming that the absence of the induction equation in the reduced set of linearized equations from Equations(\ref{eq:redu1})--(\ref{eq:redu5}) still leads to the same growth rate and wave frequency of the drag instability. Finally,
as a reminder, in the transversely large-scale regime where $k_y/k_x \ll V_n/c_s = \hat k_{jump}$ , the property of the drag instability in 2D resembles that in 1D, according to the analysis in the previous subsection. Therefore, the density growth also governs the dynamics of the drag instability for the transversely large-scale mode \citep{GC20}.

\subsection{Recap}
When $k_y/k_x \lesssim V_n/c_s \equiv \hat k_{jump}$ (i.e., transversely large-scale modes), the drag instability behaves similarly between 1D and 2D perpendicular C-shocks.  When  $k_y/k_x \gtrsim \hat k_{jump}$ (i.e., transversely small-scale modes), a new property of the growing mode emerges for 2D perpendicular C-shocks even for a large $k_x$, resulting from the ion-neutral drag coupled with the slow mode along the background magnetic field. When  $k_y/k_x \sim \hat k_{jump}$, the sign transition of wave frequency occurs. This leads to an unstable mode of a small wave frequency and thus enables a slowly propagating mode to grow substantially within a C-shock, unlimited by the shock width. The linear result suggests that the density growth dominates the evolution of the perturbation growth driven by the drag instability for the modes on both transversely  large and small scales.

\section{Linear analysis: isothermal oblique shocks}
\label{sec:oblique}

\begin{figure}
\plotone{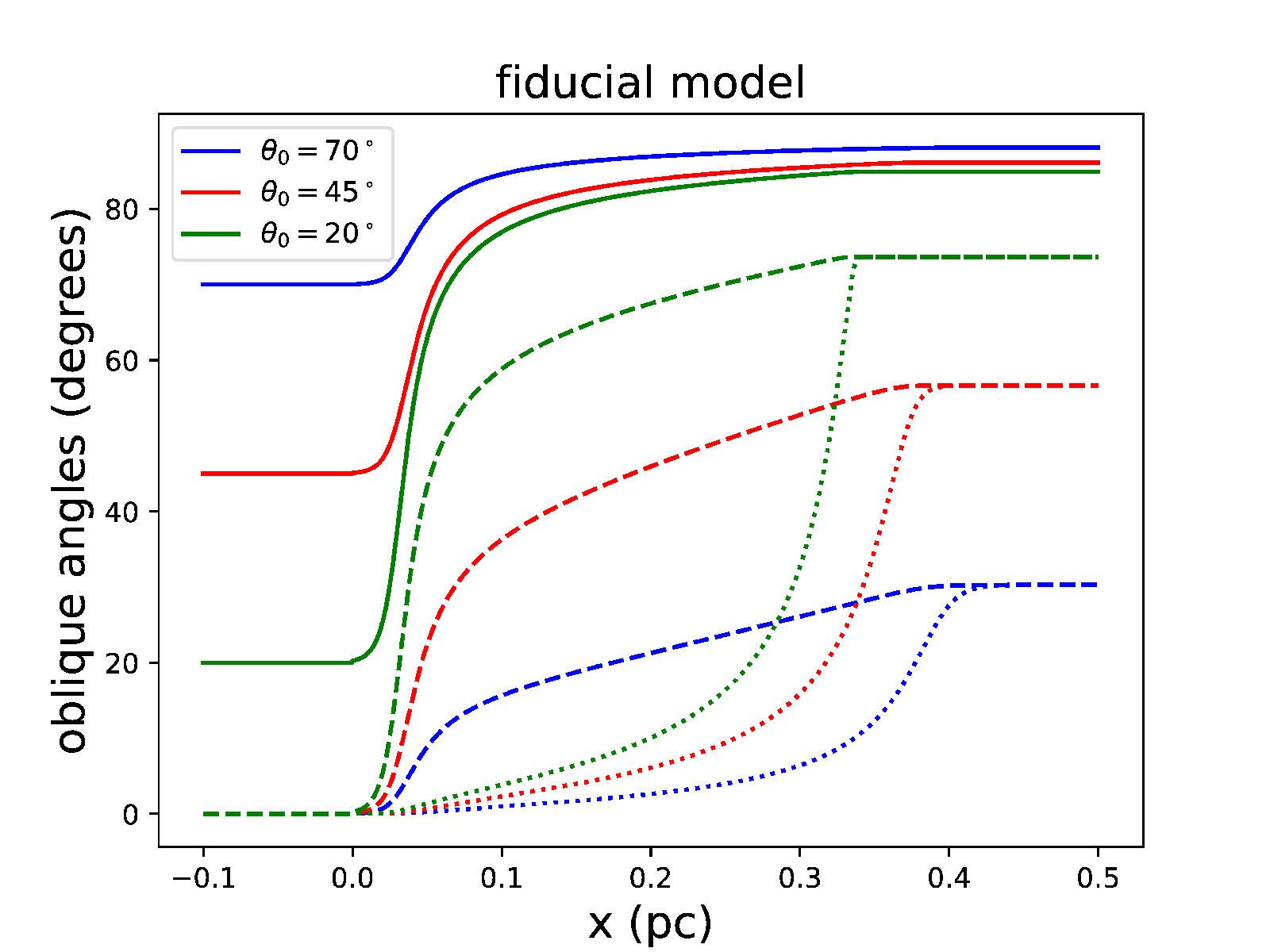}
\caption{Oblique angle of  $\bf B$ (solid curves), $\bf V_i$ (dashed curves), and $\bf V_n$ (dotted curves) relative to the direction of the pre-shock inflow (i.e., the $+x$ direction) as a function of $x$ for the fiducial model of oblique C-shocks. The cases for three different initial oblique angles $\theta_0$ (70$^\circ$, 45$^\circ$, 20$^\circ$) are presented. These results provide the background states of oblique C-shocks for the linear analysis of the drag instability.}
\label{fig:angles}
\end{figure}

\subsection{Background States and Linearized Equations}
\label{sec:bg_ob}
In this section, we study the local stability of a 2D steady oblique shock in which magnetic fields are not normal to the shock flow. We adopt the background states constructed by \citet{CO12}. That is,  the pre-shock flow is still along the $x$-direction,
the shock front in the $y$-$z$ plane, and the pre-shock magnetic field in the $x$-$y$ plane (${\bf B_0}=B_{x,0} {\hat x} + B_{y,0} {\hat y}$), at an initial oblique angle $\theta_0$ to the inflow ($B_{y,0}/B_{x,0}=\tan \theta_0$). For a steady, parallel-plane oblique shock (i.e., $\partial_t=\partial_y=\partial_z=0$), \citet{CO12} derived the equilibrium equations for $r_n \equiv \rho_n/\rho_{n,0} = v_0/V_{n,x}$, $r_B \equiv B_y/B_{y,0}$ ($B_x$ is constant because $\nabla \cdot {\bf B}=0$), and $r_{ix} \equiv v_0/V_{i,x}$ from Equations(\ref{eq:1})-(\ref{eq:5}) based on the strong coupling approximation and ionization--recombination equilibrium. Note that $r_B$ is not equal to $r_{ix}$ for an oblique shock. By solving the equilibrium equations with a set of pre-shock conditions,
the background states including $\rho_i$, $\rho_n$, $V_{i,x}$, $V_{n,x}$, $V_{i,y}$, $V_{n,y}$, and $B_y$ can be subsequently obtained as a function of $x$ within the shock and post-shock regions. Since the field line is not perpendicular to the incoming shock flow, both the field line and gas inflow will continuously change their directions relative to their initial direction as they move across the shock width until they reach the post-shock region (see Figure~\ref{fig:angles} in the case of the pre-shock conditions given by the fiducial model). As expected, $\bf B$ and $\bf V_i$ are first tilted toward the shock front (i.e. a large oblique angle shown in Figure~\ref{fig:angles}) due to the earlier compression of $\rho_i$, followed by the subsequent tilt of $\bf V_n$ toward the shock front due to the later compression of $\rho_n$ within the shock. Finally, ${\bf V_i}={\bf V_n}$ in the post-shock region as in the pre-shock region, but with a final nonzero oblique angle. The shock width is given by the region between the pre- and post-shock.
It is evident from Figure~\ref{fig:angles} that as $\theta_0$ decreases, the compression ratio increases and  shock width decreases. Besides, $\bf B$ in the three cases of $\theta_0$ shown in Figure~\ref{fig:angles} all quickly becomes more or less normal to the shock flow by shock compression. All of these results are anticipated from the analysis in \citet{CO12}.

\comment{
\begin{equation}
O U = \rmi \omega U,\label{eq:disp_matrix_oblique}
\end{equation}
where
\begin{eqnarray}
O=\left[\arraycolsep=0.01pt
\begin{array}{cccccccc}
-\rmi k_x V_{i,x}-\rmi k_y V_{i,y}-2\beta \rho_i & -\rmi k_x \rho_i  & -\rmi k_y \rho_i & 0 & 0 & \xi_\mathrm{CR} & 0 & 0\\
-\rmi k_x \frac{c^2_s}{\rho_i} & -\rmi k_x V_{i,x} -\rmi k_y  V_{i,y}-\gamma \rho_n & 0 &  \rmi k_y \frac{V^2_{A,i}}{B_y} & -\rmi k_x \frac{V^2_{A,i}}{B_y}  & -\gamma V_{d,x} & \gamma \rho_n & 0\\
-\rmi k_y \frac{c^2_s}{\rho_i} & 0 & -\rmi k_x V_{i,x}-\rmi k_y V_{i,y} -\gamma \rho_n & -\rmi k_y{B_x\over B_y }{V_{A,i}^2 \over B_y} &   -\rmi k_x{B_x\over B_y }{V_{A,i}^2 \over B_y} & -\gamma V_{d,y} & 0 & \gamma \rho_n \\
0 & \rmi k_y B_y  & -\rmi k_y B_x & -\rmi k_x V_{i,x}-\rmi k_y V_{i,y} & 0 & 0 & 0 & 0\\
0 & -\rmi k_x B _y & -\rmi k_x B_x & 0 & -\rmi k_x V_{i,x}-\rmi k_y V_{i,y} & 0 & 0 & 0\\
0 & 0 & 0  & 0 & 0 & -\rmi k_x V_{n,x}-\rmi k_y V_{n,y}  & -\rmi k_x \rho_n  & -\rmi k_y \rho_n\\
\gamma V_{d,x} & \gamma \rho_i & 0 & 0 & 0 & -\rmi k_x \frac{c_s^2}{\rho_n}  & -\rmi k_x V_{n,x}-\rmi k_y V_{n,y}-\gamma \rho_i  & 0\\
\gamma V_{d,y} & 0 & \gamma \rho_i & 0 & 0 & -\rmi k_y  \frac{c_s^2}{\rho_n}  & 0 & -\rmi k_x V_{n,x} -\rmi k_y V_{n,y}-\gamma \rho_i
\end{array}
\right].
\end{eqnarray}
}

With the background states, the linearized equations are given by
\begin{equation}
O U = \rmi \omega U,\label{eq:disp_matrix_oblique}
\end{equation}
where
\begin{eqnarray}
O=\left[\arraycolsep=0.01pt
\begin{array}{cccc}
-\rmi k_x V_{i,x}-\rmi k_y V_{i,y}-2\beta \rho_i & -\rmi k_x \rho_i  & -\rmi k_y \rho_i & 0 \\
-\rmi k_x \frac{c^2_s}{\rho_i} & -\rmi k_x V_{i,x} -\rmi k_y  V_{i,y}-\gamma \rho_n & 0 &  \rmi k_y \frac{V^2_{A,i}}{B_y} \\
-\rmi k_y \frac{c^2_s}{\rho_i} & 0 & -\rmi k_x V_{i,x}-\rmi k_y V_{i,y} -\gamma \rho_n & -\rmi k_y{B_x\over B_y }{V_{A,i}^2 \over B_y}  \\
0 & \rmi k_y B_y  & -\rmi k_y B_x & -\rmi k_x V_{i,x}-\rmi k_y V_{i,y} \\
0 & -\rmi k_x B _y & \rmi k_x B_x & 0 \\
0 & 0 & 0  & 0  \\
\gamma V_{d,x} & \gamma \rho_i & 0 & 0 \\
\gamma V_{d,y} & 0 & \gamma \rho_i & 0 
\end{array}
\right.
\nonumber
\end{eqnarray}

\begin{eqnarray}
\left.
\begin{array}{cccc}
 0 & \xi_\mathrm{CR} & 0 & 0\\
 -\rmi k_x \frac{V^2_{A,i}}{B_y}  & -\gamma V_{d,x} & \gamma \rho_n & 0\\
 \rmi k_x{B_x\over B_y }{V_{A,i}^2 \over B_y} & -\gamma V_{d,y} & 0 & \gamma \rho_n \\
 0 & 0 & 0 & 0\\
 -\rmi k_x V_{i,x}-\rmi k_y V_{i,y} & 0 & 0 & 0\\
 0 & -\rmi k_x V_{n,x}-\rmi k_y V_{n,y}  & -\rmi k_x \rho_n  & -\rmi k_y \rho_n\\
 0 & -\rmi k_x \frac{c_s^2}{\rho_n}  & -\rmi k_x V_{n,x}-\rmi k_y V_{n,y}-\gamma \rho_i  & 0\\ 
 0 & -\rmi k_y  \frac{c_s^2}{\rho_n}  & 0 & -\rmi k_x V_{n,x} -\rmi k_y V_{n,y}-\gamma \rho_i
 \end{array} 
\right],
\label{eq:M}
\end{eqnarray}
where $V_{A,i}$ is still defined as $B_y^2/4\pi \rho_i$ in terms of the field component parallel to the shock front, which follows the same definition in the case of the 2D perpendicular shock.
When the $x$-component of the pre-shock $B$ field is zero, $B_x=0$, $V_{i,y}=0$, and $V_{n,y}=0$ (thus $V_{d,y}=0$) throughout the shock. Thus, the matrix $O$ in the above equation for an oblique shock is reduced to the matrix $P$ in equation(\ref{eq:disp_matrix_2D}) for a perpendicular shock.

\begin{figure}
\plottwo{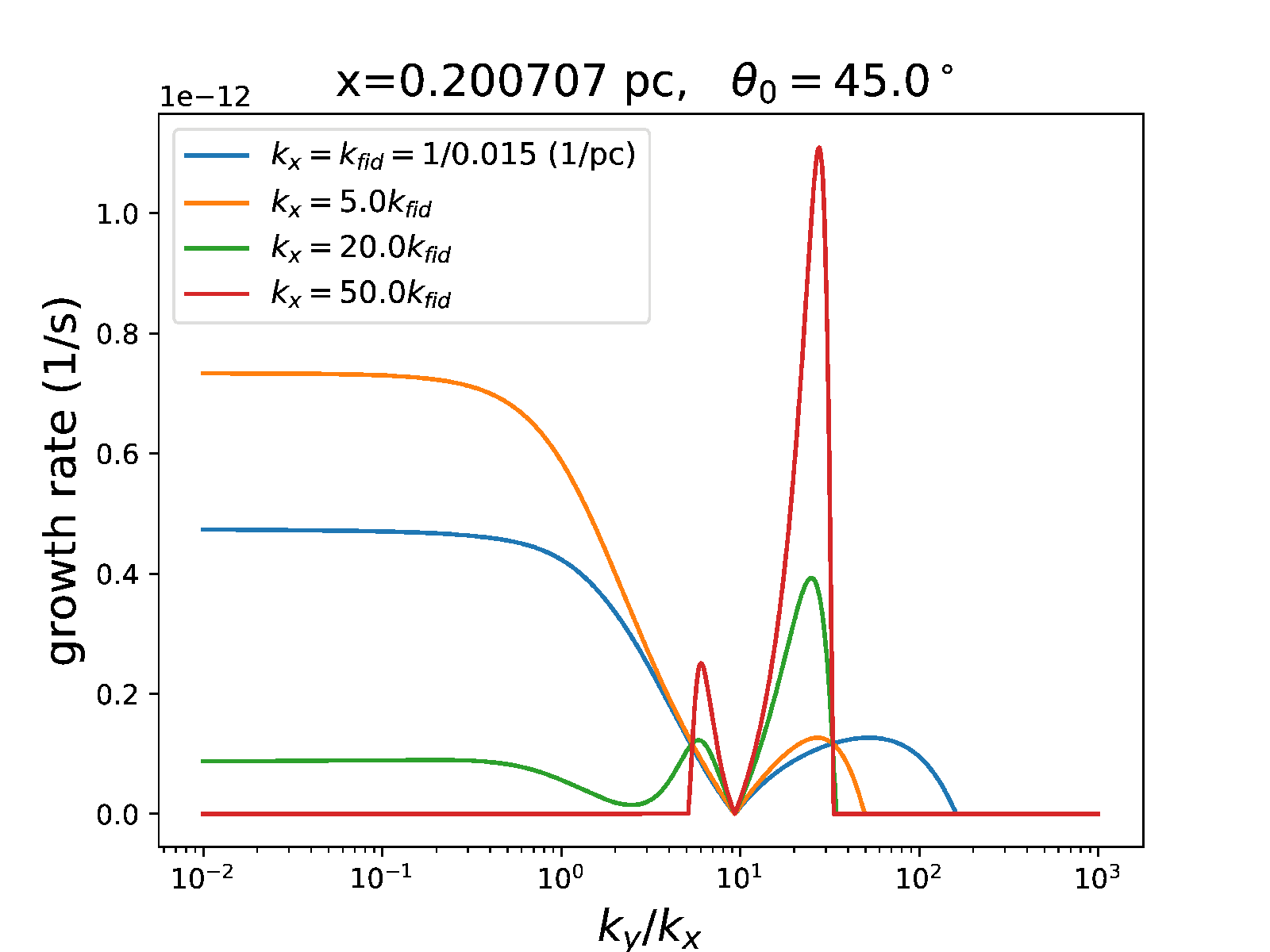}{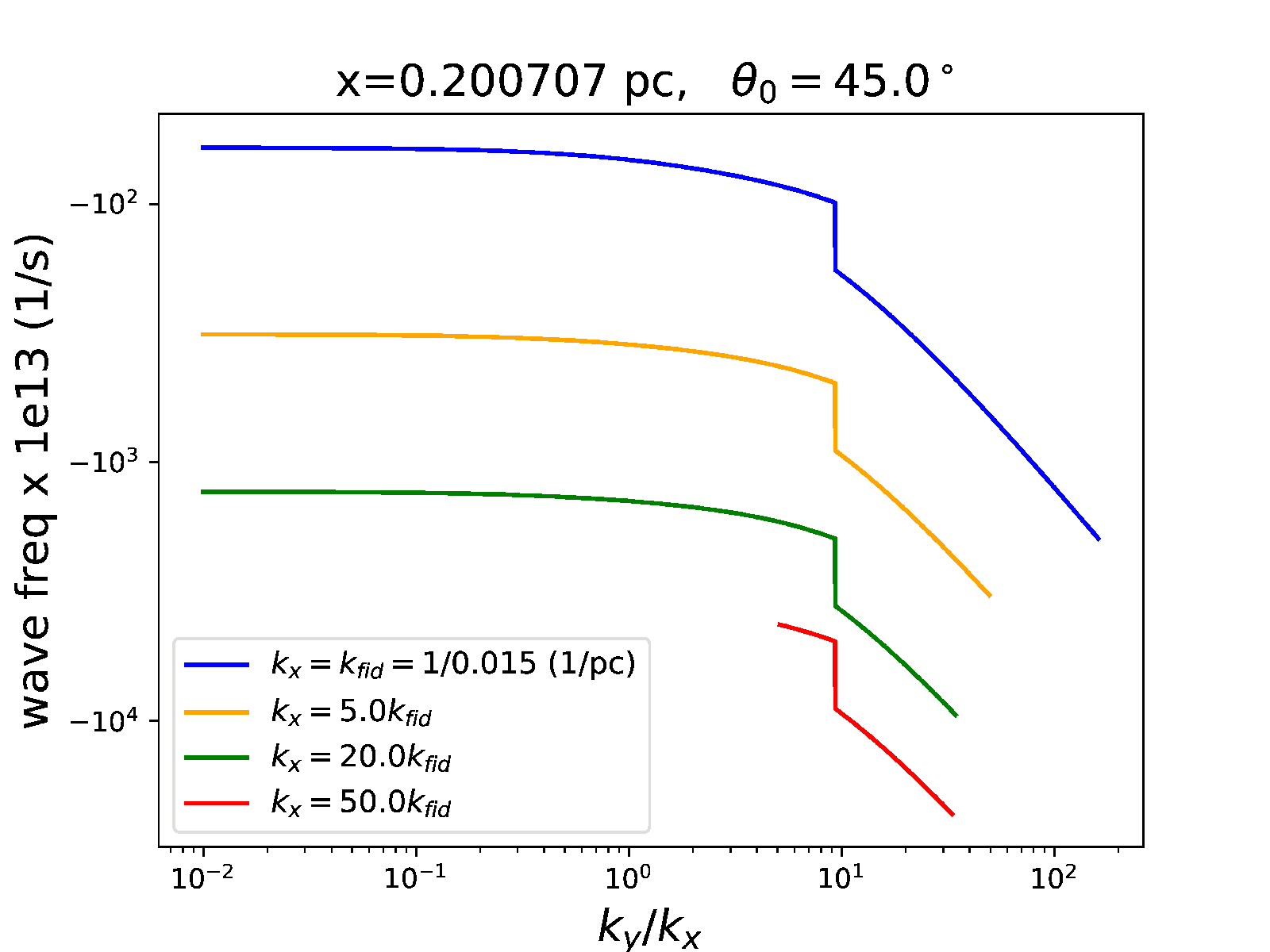}
\caption{Growth rate $\Gamma_{grow}$ (left panel) and its corresponding wave frequency in the shock frame $\omega_{wave}$ (right panel) of the drag instability in the fiducial model for a 2D oblique shock with $\theta_0=45^\circ$, plotted as a function of $k_y/k_x$ in the four cases of $k_x$ at $x\approx 0.2$ pc. The wave frequency is not presented when the growth rate is zero.}
\label{fig:mode_45}
\end{figure}

We solve the above eigenvalue problem for the fiducial model of the C-shock with a pre-shock oblique angle $\theta_0 < 90^\circ$. As in the perpendicular shock, we find that there is one unstable mode among the eight eigenmodes within a C-shock. Figure~\ref{fig:mode_45} shows the growth rate $\Gamma_{grow}$ and wave frequency in the shock frame $\omega_{wave}$ of the unstable mode as a function of $k_y/k_x$ in the case of $\theta_0=45^\circ$ with four values of $k_x$ at $x\approx 0.2$ pc, approximately in the middle of the oblique C-shock. The figure exhibits two clear general features. First of all, the growth rates and wave frequencies do not vary significantly but have values similar to the case of $k_y=0$ in the range of $k_y/k_x < 1$, as shown by the flat part of the curves. The second feature is that regardless of the difference in $k_x$, the growth rates of the unstable mode all drop to zero when $k_y/k_x \approx 9.3$, and the corresponding wave frequencies exhibit a discontinuity at the same wavenumber ratio, which turns out to be exactly equal to $|V_{d,x}|/V_{d,y}$ at $x\approx 0.2$ pc in the fiducial model. We investigate the reason in the next subsections.

\subsection{Simplified Dispersion Relation}
\label{sec:dispersion_oblique}
Since  $V_{n,y}$ and $V_{i,y}$ other than $V_{n,x}$ and $V_{i,x}$ are present in an oblique shock, the Doppler-shift rates in the comoving frame of the ions $\Gamma_i$ and neutrals $\Gamma_n$ are expressed by $\rmi (\omega + k_x V_{i,x}+k_y V_{i,y})$ and $\rmi (\omega + k_x V_{n,x}+k_y V_{n,y})$, respectively.
Following the same approach in Section~\ref{sec:dispersion} for perpendicular shocks, we focus on the particular mode with $\Gamma_n$ smaller than $2 \beta \rho_i$ and $kV_d$ but larger than $\gamma \rho_i$.
%and $\Gamma_{th}$, 
The linear equations expressed in Equation(\ref{eq:disp_matrix_oblique}) may be reduced to the same set of Equations(\ref{eq:redu3}), (\ref{eq:redu4}), (\ref{eq:redu5_1}), and (\ref{eq:redu1_1}), except that the additional drag term $\gamma V_{d,y} \delta \rho_i$ appears in the $y$-component of the momentum equation for the neutrals, namely (cf. Equation(\ref{eq:redu5_1})):
\begin{equation}
-\gamma \rho_i V_{d,y} {\delta \rho_i \over \rho_i}+ \rmi k_y c_s^2 {\delta \rho_n \over \rho_n} + \Gamma_n \delta v_{n,y} \approx 0. \label{eq:redu5_1_ob}
\end{equation}

 Moreover, as the recombination rate $2\beta \rho_i > \Gamma_n$ for this mode, Equation(\ref{eq:redu1_1}) is reduced to $\delta \rho_i/\rho_i \approx (1/2) \delta \rho_n/\rho_n$ due to the ionization equilibrium. The resulting dispersion reads
\begin{equation}
\Gamma_n^2 = -\rmi {\gamma \rho_i \over 4} (k_x V_{d,x} + k_y V_{d,y})- k^2 c_s^2.\label{eq:k_zero}
\end{equation}
In the regime where $k_y/k_x \ll 1$ and $\Gamma_n > k_x c_s$, the above equation is reduced to Equation(\ref{eq:disp_Gu}), the typical growth rate and wave frequency for the 1D drag instability \citep{Gu04,GC20}, but with a background state for an oblique shock.  As in the case of a 2D perpendicular shock discussed in the preceding section, this mode behavior of an oblique shock is consistent with the flat part of the curves for both growth rate and wave frequency in Figure~\ref{fig:mode_45}. As $k_y/k_x$ increases to the moderate value $|V_{d,x}|/V_{d,y} \equiv \hat k_{discon} =9.3$, Figure~\ref{fig:mode_45} shows that the growth rates decline and become zero around  $\hat k_{discon}$. Besides, the corresponding wave frequencies decrease gradually with increasing $k_y/k_x$ and  then abruptly drop around $\hat k_{discon}$. This behavior of the growth rate and wave frequency can be understood in terms of Equation({\ref{eq:k_zero}}), which admits a special solution Re$[\Gamma_n]=0$ when $k_x V_{d,x} + k_y V_{d,y}=0$.  The dynamics attributed to the absence of the growing mode at $k_y/k_x=\hat k_{discon}$ is that the drag force in the $x$-direction (i.e., $\gamma V_{d,x} \delta \rho_i$) acts out of phase with the drag force in the $y$-direction for the neutrals (i.e., $\gamma V_{d,y} \delta \rho_i$), therefore suppressing the density enhancement in the neutral continuity equation and thus quenching the drag instability.

To explain the property of the wave frequency shown in the right panel of Figure~\ref{fig:mode_45}, we investigate the behavior of Equation(\ref{eq:k_zero}) near $k_y/k_x=\hat k_{discon}$. Because  $k c_x \gg k_x V_{d,x}+k_y V_{d,y} \approx 0$ around this wavenumber ratio, 
the dispersion relation of the unstable mode described by Equation(\ref{eq:k_zero}) is simplified to
\begin{equation}
\Gamma_n = \rmi \omega + \rmi k_x V_{n,x} + \rmi k_y V_{n,y} \approx {\gamma \rho_i \over 4}{|k_x V_{d,x}+k_y V_{d,y}| \over k c_s}\pm \rmi k c_s,\label{eq:disp_ob}
\end{equation}
where the plus (minus) sign of the imaginary part of the above expression (i.e., the wave frequency in the comoving frame of the neutrals) is taken when $k_x V_{d,x} + k_y V_{d,y}<0$ ($>0$). Due to this sign change, the wave frequency in the shock frame $\omega_{wave}$ (=Re[$\omega$]) at $k_y/k_x=k_{discon}$ exhibits a discontinuity, with the difference given by $\approx 2kc_s$. As $k_y/k_x$ increases beyond $\hat k_{discon}$ for a given $k_x$, Re[$\omega$] ($=-k_x V_{n,x}-k_y V_{n,y}-kc_s$) increases more negatively; approximately speaking, Re[$\omega$]$\approx -k_y(V_{n,y}+c_s) \propto -k_y$, in close agreement with the right panel of Figure~\ref{fig:mode_45}.

\subsection{An auxiliary analysis based on phase differences between perturbations}

\begin{figure}
    \centering
    \subfigure[]{\includegraphics[width=0.32\textwidth]{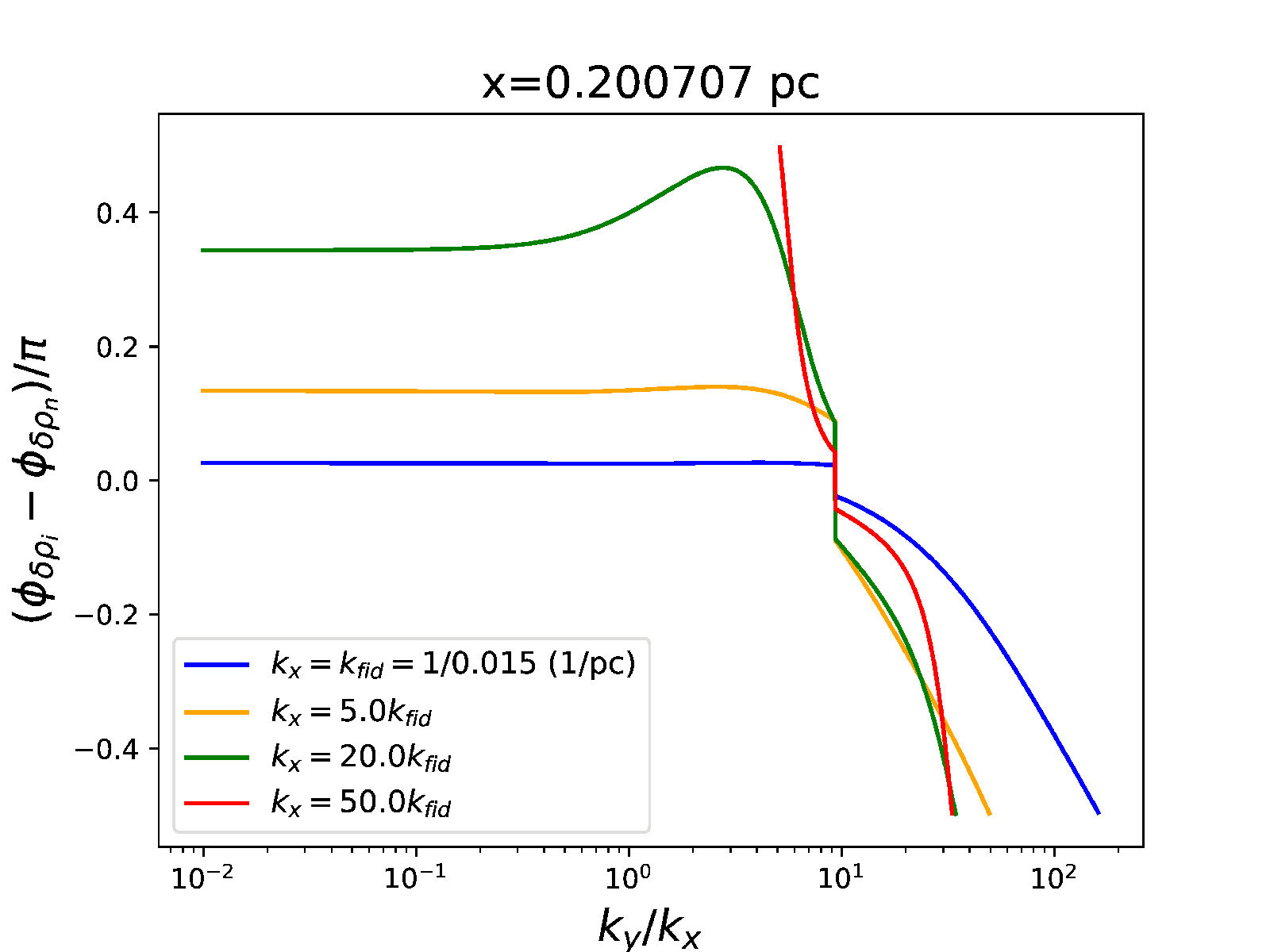}} 
    \subfigure[]{\includegraphics[width=0.32\textwidth]{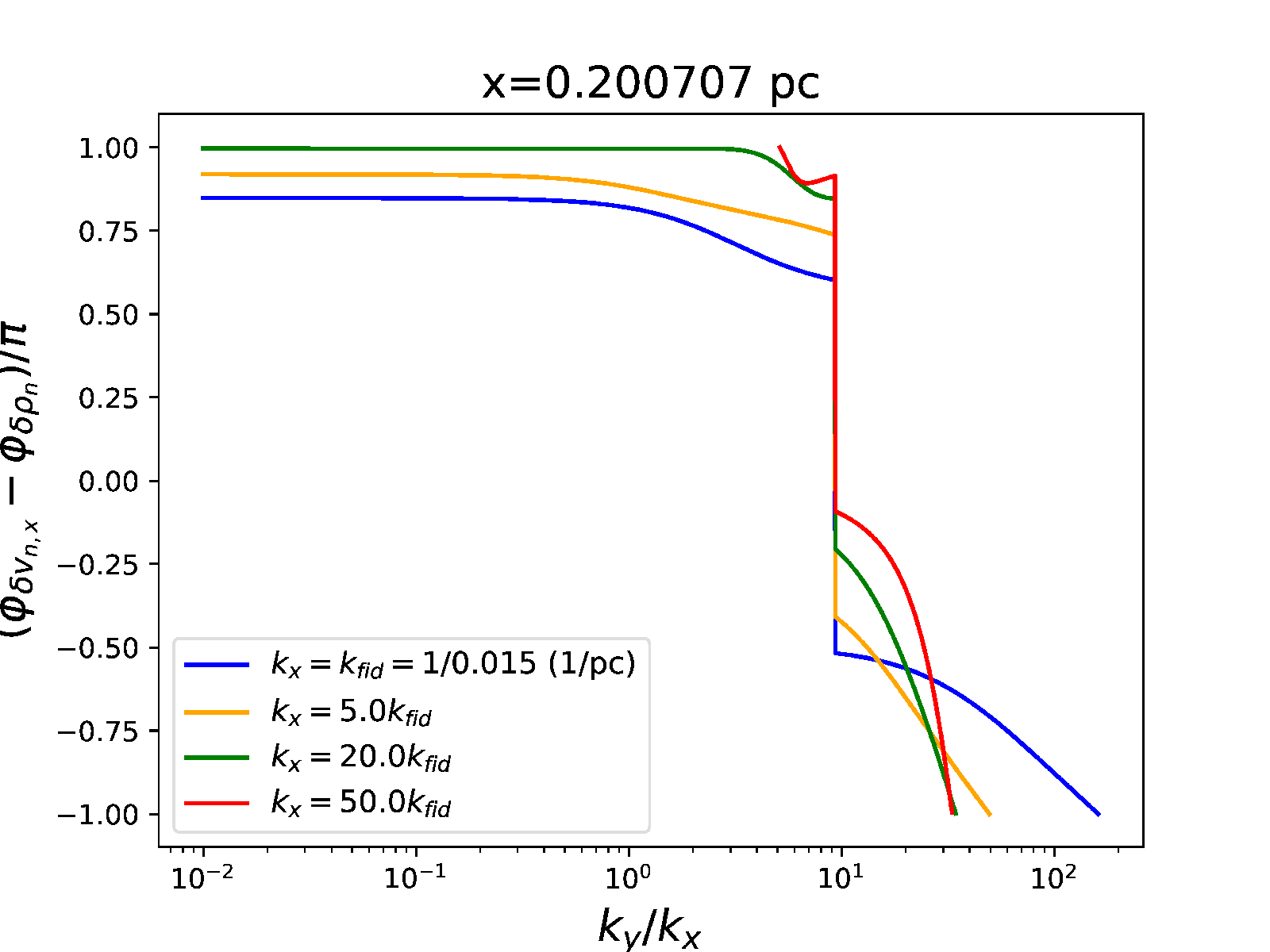}} 
    \subfigure[]{\includegraphics[width=0.32\textwidth]{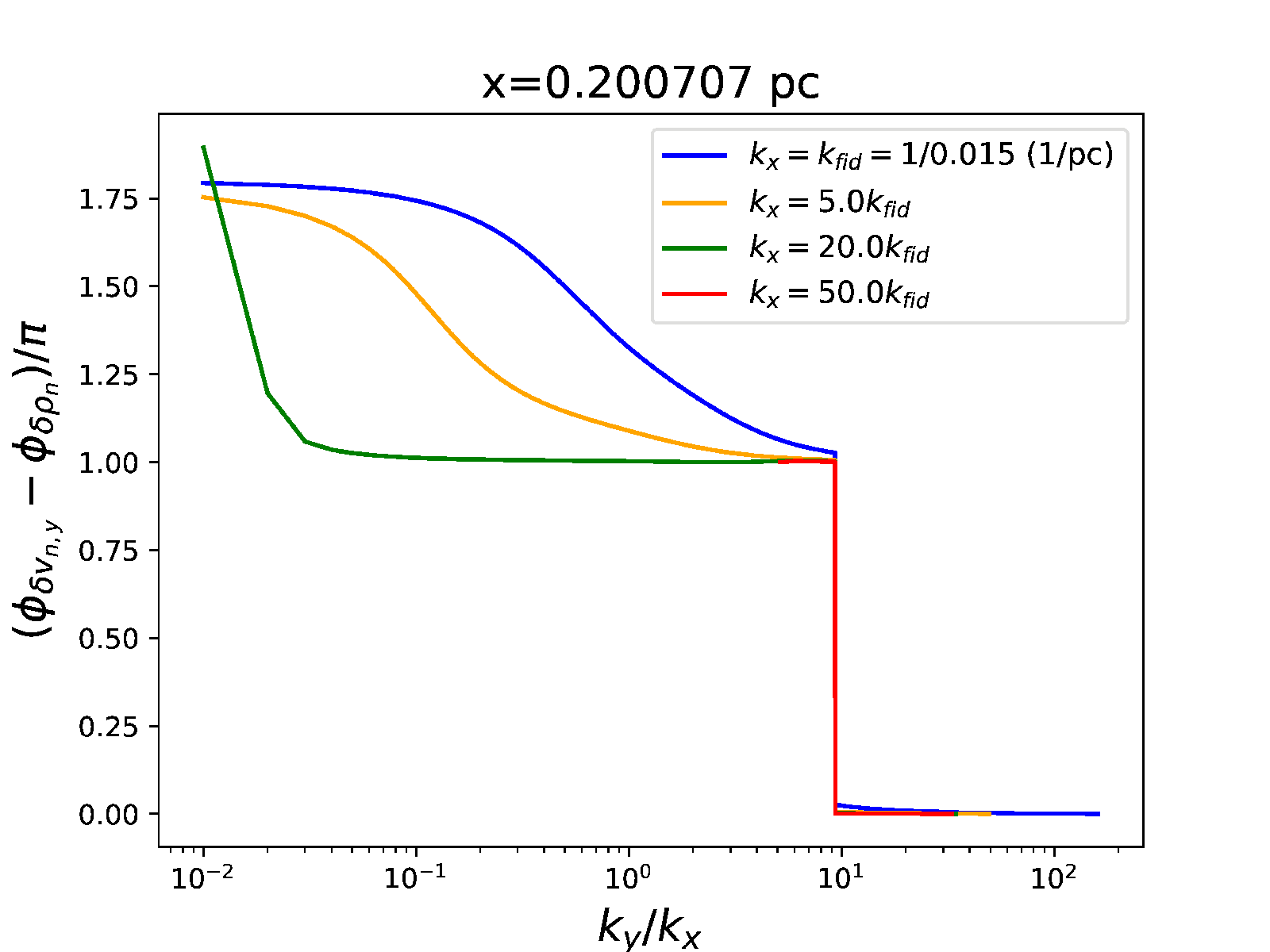}}
    \caption{Phase differences between perturbations as a function of $k_y/k_x$ for the unstable mode in the fiducial model of  an oblique shock with $\theta_0=45^\circ$. The cases of the four values of $k_x$ are shown.
    The phase differences are not presented when the growth rate is zero (refer to the left panel of Figure~\ref{fig:mode_45}).}
    \label{fig:ph_diff_ob}
\end{figure}

Although we attempt to obtain the simplified dispersion relations for guiding us in understanding the key features of the growth rate and wave frequency of an unstable wave, not all of the features can be explained. Figure~\ref{fig:mode_45} shows that the unstable mode is suppressed when $k_y/k_x$ is sufficiently large, which is unable to be described by Equation(\ref{eq:disp_ob}), because some neglected terms in deriving the simplified dispersion relations can become comparably important for a large $k_y/k_x$. To gain a more comprehensive insight into the 2D drag instability, we study the phase differences between perturbation quantities of the unstable mode as in the preceding section to complement the limited analysis from the dispersion relations. The results %corresponding to the results shown in Figure~\ref{fig:mode_45} 
as a function of $k_y/k_x$
are illustrated in Figure~\ref{fig:ph_diff_ob}. 

As in the case of perpendicular shocks, the flat part of the curves in panels (a) and (b) of Figure~\ref{fig:ph_diff_ob} correspond to that in Figure~\ref{fig:mode_45} for $k_y/k_x \lesssim 1$, arising from the fact that 2D oblique shocks resemble 1D shocks (i.e., $k_y=0$) in terms of the dynamics in the $x$-direction. In the $y$-direction, however, the phase difference between $\delta \rho_n$ and $\delta v_{n,y}$ changes gradually from $\approx 1.75 \pi$ (equivalent to $-\pi/4$) to $\pi$ as $k_y/k_x$ increases from $\approx 0.01$ to $\hat k_{discon}$ ($\approx 9.3$ at $x=0.2$ pc). It arises because when $k_y/k_x \ll 1$, the ionization equilibrium (i.e. $\delta \rho_i /\rho_i=(1/2)\delta \rho_n/ \rho_n$) and the ion-neutral drag dictate the dynamics of Equation(\ref{eq:redu5_1_ob}), yielding  the phase difference $-\pi/4$ in the $y$-direction as well. As $k_y/k_x$ increases for a given $k_x$, Equation(\ref{eq:disp_ob}) implies that the wave frequency is gradually dominated by $k c_s$ rather than the Doppler-shift frequency $k_x V_{n,x}+k_y V_{n,y}$. Additionally, the drag term becomes subdominant to the pressure term in Equation(\ref{eq:redu5_1_ob}). Thus, Equation(\ref{eq:redu5_1_ob}) leads to the relation $\delta \rho_n/\rho_n \approx -\delta v_{n,y}/c_s$; i.e., $\delta \rho_n$ and $\delta v_{n,y}$ gradually become out of phase as $k_y/k_x$ increases to $\hat k_{discon}$, as illustrated in panel (c) of Figure~\ref{fig:ph_diff_ob}. Once $k_y/k_x$ increases to $\hat k_{discon}$, the phase difference between $\delta \rho_n$ and $\delta v_{n,y}$ jumps from $\pi$ to zero because of the sign change of the wave frequency.
Furthermore, panel (c) of Figure~\ref{fig:ph_diff_ob} depicts that the phase difference between $\delta \rho_n$ and $\delta v_{n,y}$ is almost zero when $k_y/k_x \gtrsim \hat k_{discon}$. It can be realized from Equation(\ref{eq:redu5_1_ob}) with $\Gamma_n \approx -\rmi k_y c_s$ as suggested by  Equation(\ref{eq:disp_ob}) for $k_y/k_x > \hat k_{discon}$, resulting in the relation that $\delta \rho_n/\rho_n \approx \delta v_{n,y}/c_s$; i.e., $\delta \rho_n$ and $\delta v_{n,y}$ are almost in phase and are associated with an acoustic wave propagating in the $y$-direction. The resulting $v_{ph,y}$ of the acoustic wave points to the positive $y$-direction in the frame comoving with the neutrals.

Panel (a) of Figure~\ref{fig:ph_diff_ob} also shows that 
when $k_y/k_x > \hat k_{discon}$,
the phase difference between $\delta \rho_i$ and $\delta \rho_n$ increases with $k_y/k_x$, which can be realized from Equation(\ref{eq:redu1_1}). Since $\Gamma_i=\Gamma_n + \rmi (k_x V_{d,x} + k_y V_{d,y})$, the term $k_x V_{d,x}+k_y V_{d,y}$ is nearly zero at $k_y/k_x=k_{discon}$ and  thus is negligible compared to the recombination rate $2 \beta \rho_i$, resulting in the ionization--recombination equilibrium. However, as $k_y/k_x$ increases from $\hat k_{discon}$, the term $k_x V_{d,x}+k_y V_{d,y}$, which is neglected for deriving Equation(\ref{eq:k_zero}), increases and thus becomes comparable to and even larger than $2 \beta \rho_i$ at a large value of $k_y$ in the term $\Gamma_i + 2 \beta \rho_i$ of Equation(\ref{eq:redu1_1}) for the ion continuity equation. Consequently, the ionization--recombination equilibrium is poorly attained, and the phase difference between $\delta \rho_i$ and $\delta \rho_n$ shifts significantly away from zero at a large value of $k_y/k_x$.

As a result of the aforementioned phase shift between $\delta \rho_i$ and $\delta \rho_n$, the $x$-component of the linearized momentum equation for the neutrals, i.e. Equation(\ref{eq:redu4}), implies that the phase difference between $\delta \rho_n$ and $\delta v_{n,x}$ shifts accordingly away from $\sim -\pi/4$ \citep[i.e. a typical phase shift for the 1D drag instability; see][]{GC20} to a large negative value as $k_y/k_x$ increases from $\hat k_{discon}$. Together with the continuity equation for the neutrals described by Equation(\ref{eq:redu3}), the phase shift leads to the change of the growth rate with $k_y/k_x$ and yields no growth at a certain large value of $k_y/k_x$ ($\sim 10^2$) where $\delta \rho_n$ and $\delta v_{n,x}$ are almost out of the phase, as shown in panel (b) of Figure~\ref{fig:ph_diff_ob}. At this point, the wave is no longer unstable but is transformed into an acoustic wave, with $v_{ph,x}$ pointing to the negative $x$-direction in the comoving frame of the neutrals. The $v_{ph,y}$ of the acoustic wave points to the positive $y$-direction in the comoving frame with the neutrals, as was explained earlier in this subsection.

\subsection{\bf Growth of an unstable mode and shock obliquity}

\begin{figure}
\plottwo{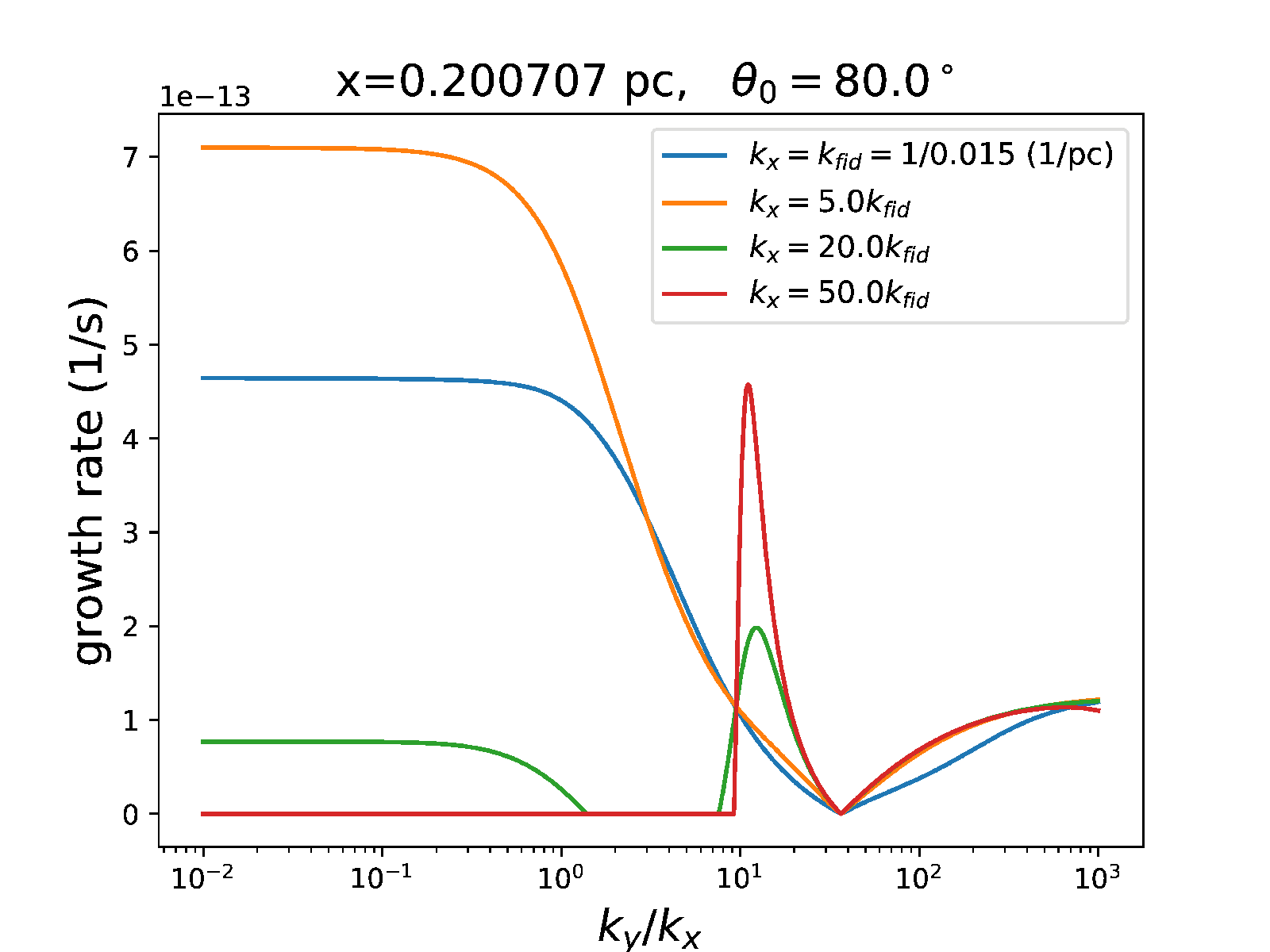}{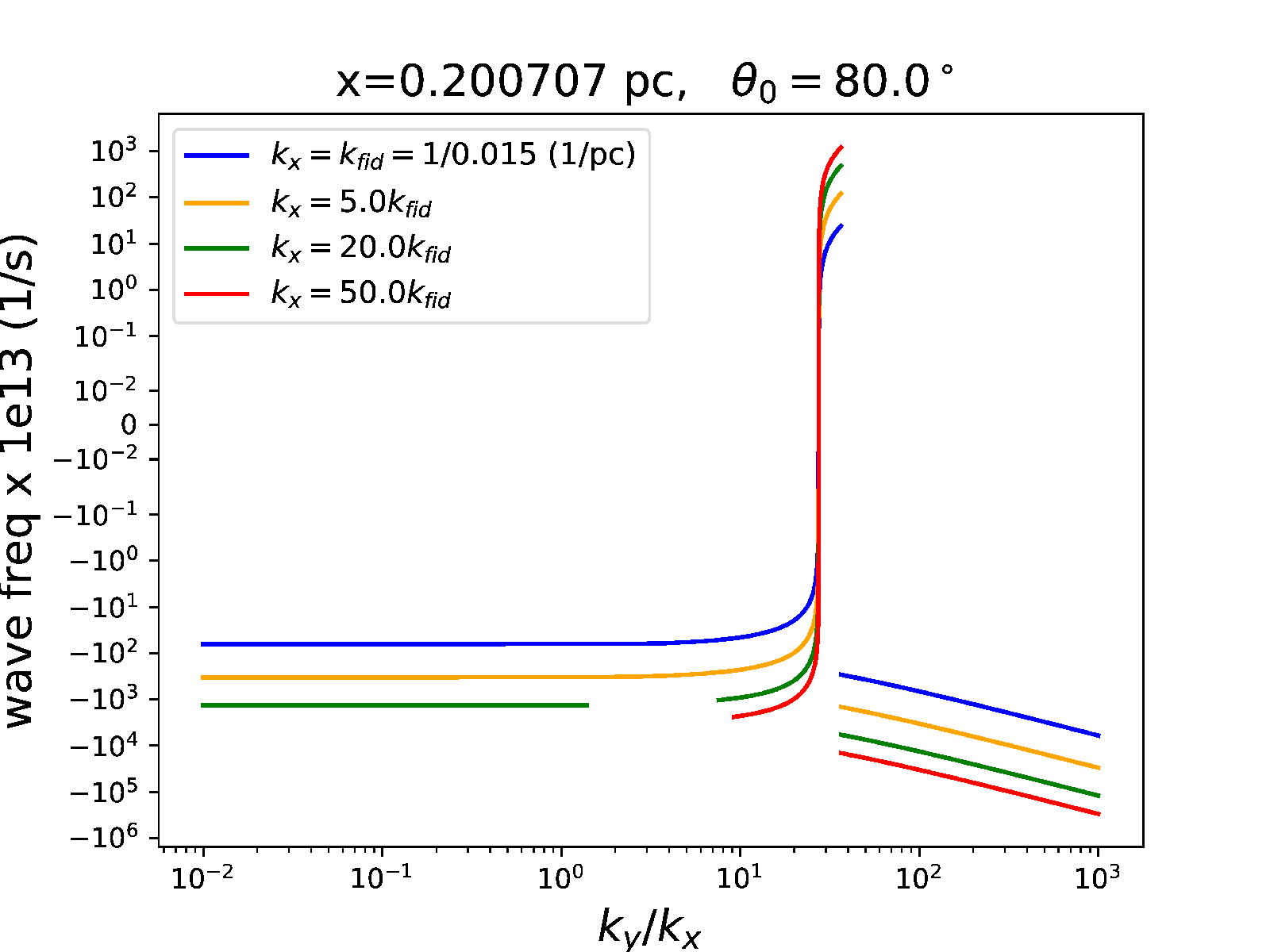}
\caption{Same as Figures~\ref{fig:ps} \& \ref{fig:mode_45} but for $\theta=80^\circ$.
The wave frequency is not presented when the growth rate is zero.}
\label{fig:mode_80}
\end{figure}

The distinct behaviors of wave frequency for a large value of $k_y/k_x$ -- frequency jump versus discontinuity -- between Figure~\ref{fig:ps} (perpendicular shock with $\theta_0=90^\circ$) and Figure~\ref{fig:mode_45} (oblique shock with $\theta_0=45^\circ$) suggests that a shock with a proper range of the oblique angle between $45^\circ$ and $90^\circ$ can
allow for the unstable modes with both behaviors of wave frequency. Figure~\ref{fig:mode_80} shows the results for the same shock model with $\theta_0=80^\circ$. Indeed, the right panel of the figure shows that a frequency jump from negative to positive values happens at $k_y/k_x \approx V_{n,x}/(c_s-V_{n,y}) \equiv \hat k_{jump} \approx 27$, and a frequency discontinuity occurs at an even smaller scale where $k_y/k_x=|V_{d,x}|/V_{d,y} \equiv \hat k_{discon} \approx 37$. Consequently, the growth rate of the modes with large $k_x$ (i.e., $20k_{fid}$ and $50k_{fid}$) rises around $\hat k_{jump}$ and the growth rates all go to zero at $\hat k_{discon}$, as shown and expected in the left panel of Figure~\ref{fig:mode_80}.
Note that $\hat k_{jump}$ involves $V_{n,y}$ for an oblique shock because the Doppler-shift frequency is $k_x V_{n,x} + k_y V_{n,y}$. 
%with a non-vanishing $V_{n,y}$.
What happens is that as the oblique angle $\theta_0$ decreases from $90^\circ$, $V_{n,y}$ and thus $V_{d,y}$ start deviating from zero such that $\hat k_{jump}<\hat k_{discon}<\infty$. Consequently, the modes with behaviors of the frequency jump and discontinuity both appear in the case for $\theta_0=80^\circ$. As $\theta_0$ continues to decrease (i.e. the shock is more oblique) until $\hat k_{discon}$ becomes smaller than $\hat k_{jump}$, the mode with the behavior of a frequency jump disappears at $k_y/k_x=\hat k_{jump}$, leaving the sole phenomenon of the frequency discontinuity at $k_y/k_x=\hat k_{discon}$ such as the case for $\theta_0=45^\circ$. At the location of $x\approx 0.2$ pc in the fiducial model, $\hat k_{jump}=\hat k_{discon}$ occurs between $\theta_0=78^\circ$ and 79$^\circ$; i.e., it occurs when the shock is mildly oblique. Analogous to perpendicular shocks, the presence of $\hat k_{jump}$ enables substantial growth of the slowly propagating unstable mode (i.e., unlimited by the shock width) in a mildly oblique shock. Consequently, the range of  $\hat k_{jump}$ is similar to that for the perpendicular shocks in the typical environments of star-forming clouds.

Because the slowly traveling mode with substantial growth appears only in a background environment close to a perpendicular shock, the growth of the drag instability in a moderately or exceedingly oblique shock is still limited by the short time span for the unstable mode to remain within a shock. We explore this issue by comparing the C-shock model described by the fiducial model and by the model V06 from Table 2 in \citet{GC20}. The pre-shock conditions of model V06 are given by $n_0=200$ cm$^{-3}$, $v_0=6$ km/s, $B_0=10\mu$G, and $\chi_{i0}=5$. 
Model V06 is discussed in \citet{GC20} because it exhibits a larger MTG than that of the fiducial model in a 1D shock due to its broader shock width. 
The MTG is easily computed for a 1D shock, whereas the same calculation involves a more elaborate work for a 2D shock. In general, the growth rate $\Gamma_{grow}$ of a mode with a particular wave frequency $\omega_{wave}$ is a function of both $k_x$ and $k_y$, which vary with $x$ in a 2D shock. In this work, we do not intend to perform an accurate analysis to identify the combination of $k_x(x)$ and $k_y(x)$ that provides the MTG of the unstable mode. Rather, 
 we keep $k_y$ uniform but allow $k_x$ to change with $x$ across a shock for a given wave frequency of an unstable mode. We then estimate the MTG of an oblique shock by varying the wave frequency for a few constant $k_y$, which we refer to as MTG$_{k_y}$ in short.
Comparing MTG$_{k_y}$ for different $k_y$ and $\theta_0$ would provide the guidance on how the MTG varies with the initial oblique angle $\theta_0$. The results are enumerated in Table~\ref{tab:MTG_fid} for the fiducial model and Table~\ref{tab:MTG_V06} for model V06. Given the $k_y$ of an unstable mode and $\theta_0$ for each C-shock model, the tables display MTG$_{k_y}$ with the corresponding mode frequency $\omega_{wave}$ as well as the value of $k_x$ (in terms of $k_y/k_x$) ranging from the beginning to the end of the C-shock width.

Table~\ref{tab:MTG_fid} shows that in the fiducial shock model, MTG$_{k_y}$ does not vary significantly with $\theta_0$. The smaller the $\theta_0$, the larger the compression of the density and the magnetic field (see Figure~\ref{fig:angles}), leading to a slightly larger growth rate. However,  a smaller $\theta_0$ results in a slightly narrower shock width  (see Figure~\ref{fig:angles} and the parameter $L_{shock}$ in Table~\ref{tab:MTG_fid}). As a result of the competition between these two moderate effects, the oblique angle does not noticeably affect the overall growth of the drag instability in the fiducial model. Moreover,
when $k_y/k_{fid}  \lesssim 1$,  MTG$_{k_y}$ is comparable to the MTG for the 1D shock \citep[$= 9.9$; refer to][]{GC20} and is smaller by more than a factor of 2 for $k_y/k_{kid}=10$. It can be expected from the range of the resulting $k_y/k_x$ shown in Table~\ref{tab:MTG_fid} . When $k_y/k_x  \ll 1$, the instability behaves as a 1D mode for which $k_y=0$. On the other hand, when $k_y/k_x \gtrsim 1$, the overall growth rate decreases within the shock, and thus MTG$_{k_y}$ is small. 

\begin{figure}
\plottwo{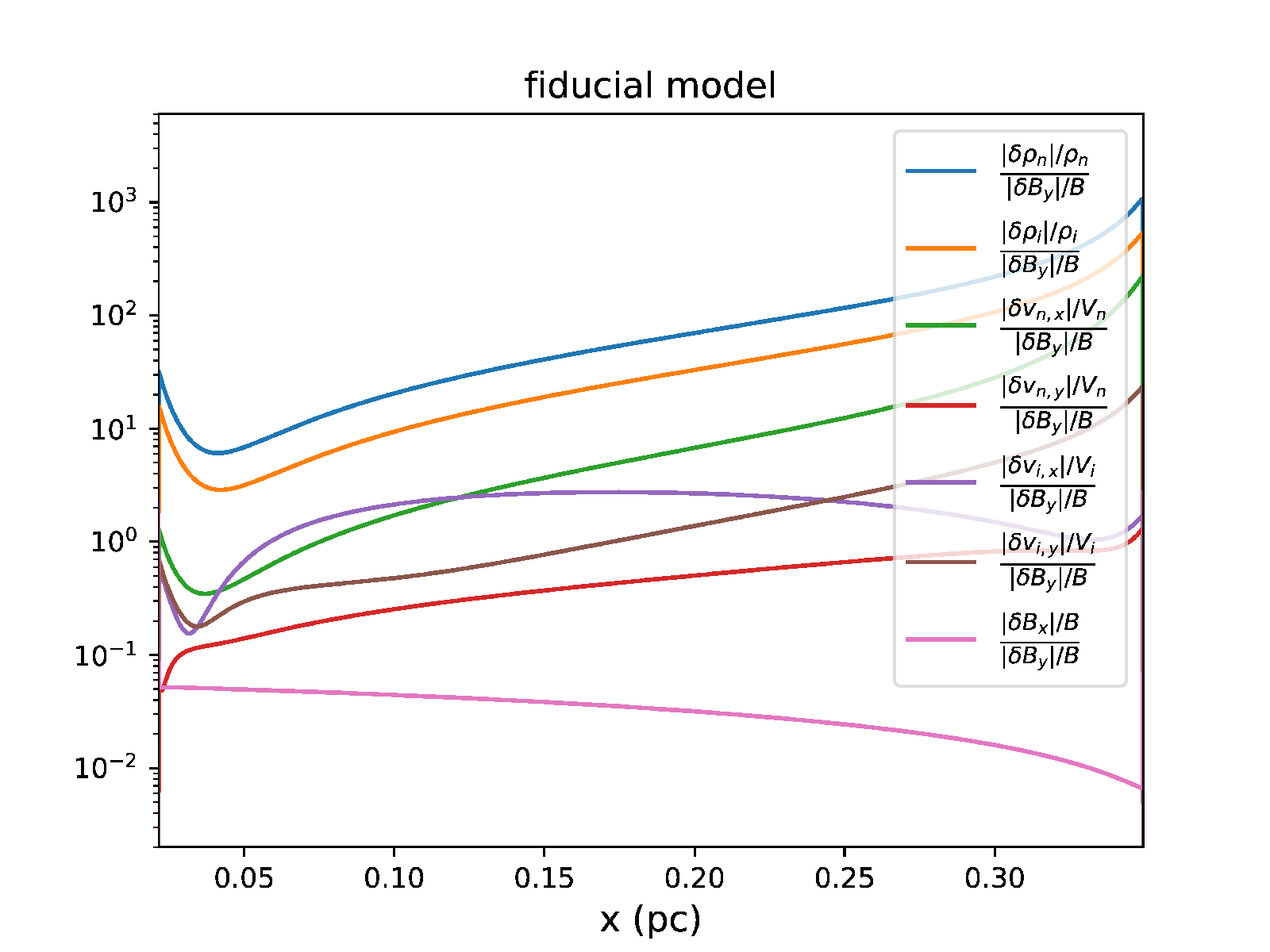}{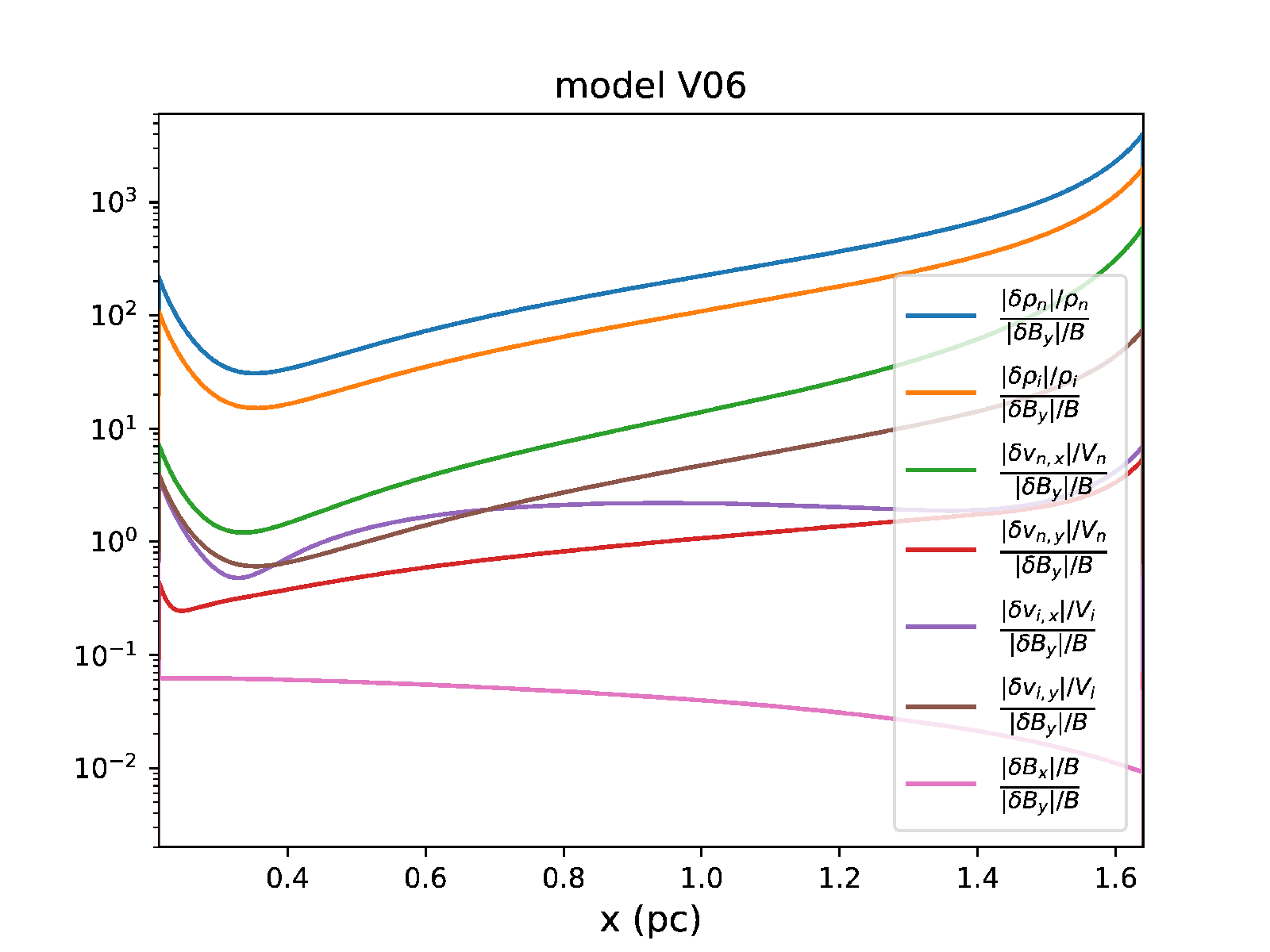}
\caption{The amplitude of perturbations normalized by the $y$ component of the magnetic perturbation $|\delta B_y|/B$ for the unstable mode with $k_y/k_{fid}=0.1$ and $\omega_{wave}=-2 \times 10^{-11}$~s$^{-1}$ in the fiducial model (left panel) and model V06 (right panel) of the C-shock with $\theta_0=45^\circ$.}
\label{fig:eigen_45}
\end{figure}

Similar to the fiducial model, Table~\ref{tab:MTG_V06}  shows that MTG$_{k_y}$ in model V06 does not vary significantly with $\theta_0$ as well. Though in the case of $k_y/k_{fid}=0.1$,  MTG$_{k_y}$ can increase moderately from 36 for $\theta_0=70^\circ$ to 47 for $\theta_0=20^\circ$. This trend actually also happens in the fiducial model with $k_y/k_{fid}=0.1$ but is less prominent (see  Table~\ref{tab:MTG_fid}).  For a smaller $\theta_0$, the effect of the slightly larger growth rate predominates more than the effect of the slightly narrower width, resulting in a larger MTG$_{k_y}$.
Evidently, a wider shock in model V06 (see the parameter $L_{shock}$ in Tables~\ref{tab:MTG_fid} and \ref{tab:MTG_V06}) allows an unstable mode to grow more and therefore makes this trend clearer.
In contrast, the two competing effects are comparable for $k_y/k_{fid}=1$ and 10; hence, the MTG$_{k_y}$ changes less noticeably with $\theta_0$.  
Similar to the fiducial model, the overall trend of the decrease in MTG$_{k_y}$ with increasing $k_y$ also happens in model V06, for the same reason.
Tables~\ref{tab:MTG_fid} and \ref{tab:MTG_V06} also show that MTG$_{k_y}$ in model V06 is in general larger than that in the fiducial model, which is expected because the shock in model V06 has a broader shock width for the instability to acquire more time to grow.
In summary, for a C-shock with the oblique angle $\lesssim 70^\circ$, we expect that the MTG of the drag instability in the fiducial model is about 10 and that in model V06 is about 30-47. Therefore, we expect that the unstable mode responsible for the MTG has $k_y/k_x \lesssim 1$. 

Figure~\ref{fig:eigen_45} shows the magnitude of the perturbations across the shock width for the unstable mode corresponding to $k_y/k_{fid}=0.1$ and $\theta_0=45^\circ$ in Tables~\ref{tab:MTG_fid} (left panel) and \ref{tab:MTG_V06} (right panel). It is evident from the figure that the magnitude of the density perturbation is always larger than that of the velocity and magnetic field perturbations of the unstable mode. In fact, it is true for all cases listed in Tables~\ref{tab:MTG_fid} and \ref{tab:MTG_V06}. The density enhancement is expected to play a critical role in the dynamics of the drag instability in an oblique shock as well.

When $\theta_0 \lesssim 20^\circ$ in our fiducial model, $B_y$ and $\rho_n$ are compressed extremely quickly near the beginning and the end of the C-shock width, respectively. Therefore, the WKBJ approximation with $k_x=k_{fid}$ becomes invalid in these shock regions. Except for these regions, the basic behavior of the drag instability for $\theta_0 \lesssim 20^\circ$ is similar to that for $\theta_0 =45^\circ$, shown in Figure~\ref{fig:mode_45} at $x\approx 0.2$ pc. When $\theta_0$ is smaller than the critical angle  $\theta_{crit} \approx 6^\circ$, the background state admits two additional solutions with field reversal, referred to as intermediate shocks \citep[e.g.,][]{Wardle1998,CO14}. In this study, we restrict ourselves to the oblique shocks without field reversal and leave the instability analysis for intermediate shocks to a future work.

\begin{deluxetable*}{cccccc}
%	\tablenum{2}
	\tablecaption{MTG$_{k_y}$ for various $k_y$ and $\theta_0$ in the fiducial model \label{tab:MTG_fid}}
	\tablewidth{0pt}
	\tablehead{
		\colhead{$k_y/k_{fid}$} & \colhead{$\theta_0$}  & $L_{shock}$ (pc) & \colhead{$\omega_{wave}$ (1/s)} &
		\colhead{$k_y/k_x$}  &\colhead{MTG$_{k_y}$}
	}
	\startdata
	0.1 & 70$^\circ$ & 0.48 & $-3$e$-11$ & 0.035--0.005 & 10.0   \\
	0.1 & 45$^\circ$ & 0.40 & $-2$e$-11$ & 0.052--0.007 & 10.3 \\
	0.1 & 20$^\circ$ & 0.24 & $-2$e$-11$ &  0.052--0.005 & 10.9 \\
	1 & 70$^\circ$ & 0.48  & $-3$e$-11$ & 0.35--0.05 & 9.57  \\
	1 & 45$^\circ$ & 0.40 & $-3$e$-11$ & 0.35--0.05  & 9.49 \\
	1 & 20$^\circ$ & 0.24& $-3$e$-11$ & 0.35--0.05  & 9.55  \\
	10 & 70$^\circ$ &  0.48  &$-5$e$-11$  & 1.97--0.32 & 4.4  \\
	10 & 45$^\circ$ & 0.40  & $-5$e$-11$  & 2.4--0.32 & 3.8  \\
	10 & 20$^\circ$ & 0.24 & $-5$e$-11$ & 2.4--0.08 & 3.4 \\
	\enddata
	\tablecomments{Given a uniform value of $k_y$ (in units of $k_{fid}$), the mode frequency $\omega_{wave}$ and the corresponding range of $k_x(x)$ (presented in terms of $k_y/k_x$) associated with the MTG at a constant $k_y$ (i.e., MTG$_{k_y}$) of the drag instability are listed for the fiducial model of a steady C-shock with various oblique angles $\theta_0$.  The shock width $L_{shock}$ is estimated using Equation(A19) in \citet{CO12}. }

\end{deluxetable*}

\begin{deluxetable*}{cccccc}
	\tablecaption{Same as Table~\ref{tab:MTG_V06} but for the C-shock model given by model V06.  \label{tab:MTG_V06}}
	\tablewidth{0pt}
	\tablehead{
		\colhead{$k_y/k_{fid}$} & \colhead{$\theta_0$}  & $L_{shock}$ (pc) & \colhead{$\omega_{wave}$ (1/s)} &
		\colhead{$k_y/k_x$}  &\colhead{MTG$_{k_y}$}
	}
	\startdata
	0.1 & 70$^\circ$ &2.08  & $-2$e$-11$ & 0.062--0.011 & 36   \\
	0.1 & 45$^\circ$ &1.70 & $-2$e$-11$ &  0.062--0.009 & 40 \\
	0.1 & 20$^\circ$ & 1.0 &$-2$e$-11$ &  0.063--0.007 & 47 \\
	1 & 70$^\circ$ & 2.08 & $-3$e$-11$ & 0.42--0.07 & 31  \\
	1 & 45$^\circ$ & 1.70 & $-3$e$-11$ & 0.42--0.07  & 30 \\
	1 & 20$^\circ$ & 1.0& $-3$e$-11$ & 0.45--0.05  & 31  \\
	10 & 70$^\circ$ & 2.08& $-7$e$-11$  & 1.72--0.33 & 7.3  \\
	10 & 45$^\circ$ & 1.70 &$-8$e$-11$  & 1.75--0.28 & 5.9  \\
	10 & 20$^\circ$ & 1.0& $-8$e$-11$ & 1.73--0.24 & 5.7 \\
	\enddata
\end{deluxetable*}

\section{Summary and Discussions}
\label{sec:sum}
In this work, we extend the study of the drag instability in 1D perpendicular C-shocks by \citet{GC20}
to 2D perpendicular and oblique C-shocks. 
We focus on the fiducial model for an isothermal steady C-shock with the pre-shock conditions described by the same fiducial model as  \citet{GC20} in the typical environment of star-forming clouds.
The WKBJ linear analyses are subsequently performed based on the background states in the fiducial model. To understand the underlying physics for the linear results, we make an attempt to derive simplified dispersion relations, aided by the auxiliary analysis of phase differences between perturbation quantities.
We observe that the drag instability remains in a 2D shock, and its behavior in general depends on $k_y/k_x$. When $k_y/k_x \lesssim 1$ (i.e., transversely  large-scale modes), the growth rate $\Gamma_{grow}$ and wave frequency $\omega_{wave}$ of the drag instability in a 2D shock are similar to those in a 1D shock, which is insensitive to the initial oblique angle $\theta_0$ of the shock. When $k_y/k_x \gtrsim 1$ (i.e., transversely  small-scale modes), the drag instability is characterized by an unstable mode coupled with the acoustic mode primarily along the $y$-direction (note that the acoustic mode is the slow mode in the case of a perpendicular shock). Additionally, in contrast to the perpendicular shock, $V_{d,y}$ exists in an oblique shock. Therefore, there exists a particular mode for an oblique shock with $k_y/k_x = 
\hat k_{discon}$ where the growth rate is zero and discontinuity in the wave frequency appears (see Equation(\ref{eq:disp_ob})).

When the shock is less oblique (i.e., $\theta_0 \gtrsim 80^\circ$ in the fiducial model), there exists a jump transition of wave frequency in the shock frame -- from the negative Doppler-shift frequency due to the shock flow to the positive acoustic wave frequency in the $y$-direction -- for a mode with $k_y/k_x \sim \hat k_{jump}$. 
Owing to the small wave frequency in the shock frame, this unstable mode propagates slowly within a shock and thus has sufficient time to potentially grow to a nonlinear phase, thereby contributing to the maximum growth. While the density enhancement of the ions approximately lies in phase with that of the neutrals by means of the ionization equilibrium for $k_y/k_x < \hat k_{jump}$,  the same phase overlap for the density perturbations is maintained by the fast acoustic wave for $k_y/k_x > \hat k_{jump}$. For the mode with an exceedingly large $k_x$, there is no growing mode for $k_y/k_x \lesssim 1$, as the drag instability is suppressed by the pressure effect \citep{Gu04,GC20}. However, unstable modes appear for $k_y/k_x \sim \hat k_{jump}$, as $k_x$ is large enough for the Doppler-shift frequency of the slowly traveling wave to dominate over the ionization rate, which in turn produces a proper phase difference between $\delta v_{n,x}$ and $\delta \rho_n$ for the drag instability to occur.

On the other hand, when the shock is more oblique (i.e., $\theta_0 \lesssim 80^\circ$ in the fiducial model), this slowly propagating unstable mode disappears. The maximum growth of the drag instability is limited by the short time span of an unstable mode to stay within a shock and hence is given by MTG, as is the case for a 1D perpendicular shock. We compute $MTG_{k_y}$, i.e. the MTG for a constant $k_y$, to infer the MTG of the drag instability for a given $\theta_0$.
We find that the MTG of the drag instability is contributed by the mode with $k_y/k_x \ll 1$ (i.e., almost 1D mode) and is expected to be about 10 insensitive to the initial oblique angle $\theta_0$ in the fiducial model. We also conduct the linear analysis for the C-shock model V06, which has a larger shock width than that of the fiducial model and thus exhibits a larger MTG in the 1D shock \citep{GC20}. We find that the MTG of the drag instability arises from the mode with $k_y/k_x \ll 1$ as well and increases from about 36 to 47 as the initial oblique angle $\theta_0$ decreases from 70$^\circ$ to 20$^\circ$. The overall larger MTG in model V06 is expected for a shock with a larger width. A larger MTG for a smaller $\theta_0$ is primarily caused by the stronger shock compression. In all the cases that we consider (see Tables~\ref{tab:MTG_fid} and \ref{tab:MTG_V06}, as well as the case of the perpendicular shock), the magnitude of the density perturbations is much larger than that of the velocity and magnetic field perturbations (e.g., Figures~\ref{fig:eigenv} \& \ref{fig:eigen_45}), implying that the density enhancement predominantly governs the dynamics of the instability in the linear regime. 

Self-gravity is not considered in our linear analysis in order to study and present the basic properties of the drag instability in 
a clean physical picture. The background magnetic fields with different initial oblique angles tend to be approximately parallel to the shock front during shock compression (see Figure~\ref{fig:angles}). In this work,
the minimal value of $k_y/k_{fid}$ that we show for the results  is  0.01. This transversely  large-scale mode has a transverse  wavelength larger than the Jeans scale ($\sim c_s/\sqrt{G \rho_n}$) and therefore may be subject to the gravitational instability along the compressed field lines primarily in the $y$-direction. However, it should be kept in mind that the background state of the shock is not static; thus, applying the Jeans criterion is debatable. 
%marginally avoid the gravitational instability of shocked gas along the background field lines primarily in the $y$ direction. 
Our analysis based on $MTG_{k_y}$  suggests that the transversely  large-scale mode would give the MTG of the drag instability within a shock (see Tables~\ref{tab:MTG_fid} \& \ref{tab:MTG_V06}),\footnote{The transversely  large-scale mode with $k_y/k_{fid}=0.1$ shown in the tables is marginally gravitationally stable against thermal pressure along the field direction almost in the $y$-direction due to shock compression. Having said that, we should bear in mind that the shock is not a static structure for the Jeans criterion to reasonably apply.} except when the shock is mildly oblique. 
Considered together, after the most unstable mode of the drag instability sets in, it
could grow to become gravitationally unstable more quickly in the direction of the field line against thermal pressure as a result of its larger transverse  scale. In this regard for a laminar background, it could be interesting to investigate the interaction between the drag and gravitational instabilities by including self-gravity in the linear analysis.

While we assume that C-shocks arise from supersonic, turbulent flows, the effects of turbulence 
are not actually modeled in our linear analysis using any phenomenological approaches, such as turbulent diffusion or turbulent pressure. The turbulent diffusion, if it exists, may weaken/eliminate transversely  small-scale modes of the drag instability or may promote the drag instability by enhancing ambipolar diffusion in a shock \citep[e.g.,][]{LN04}. The supersonic turbulent pressure, if it exists, may support the shocked gas against gravitational collapse along the compressed field lines on the scales smaller than the turbulent Jeans scale. On the other hand, supersonic turbulence in shocks is known to dissipate quickly over one turbulent crossing time, provided that there is no energy supply from the turbulent injection scale by feedback processes from protostars \citep[e.g.,][]{MacLow98,Stone98,MacLow99,NL05}. Nevertheless, even large-scale modes within C-shocks would not be affected by the decay of the large-scale turbulence, as that would only reduce the strength of C-shocks present in the environment, rather than modifying the behavior of an individual C-shock during its passage. In a nearly perpendicular shock, the slowly propagating modes responsible for the maximum growth are transversely  small-scale modes (i.e., $k_y/k_x \sim \hat k_{jump} \sim 10$ in our models).  Analogous to dense regions seeded by small-scale turbulence in a shock-compressed layer \citep{CO14},  the small-scale turbulence may help initiate the perturbation of these small-scale modes of the drag instability in compressed gas within shocks. As the modes grow to the nonlinear regime, the nonlinear saturation of the drag instability by itself could be the prime candidate to drive the turbulence within a C-shock. Compression of background turbulence is likely to be negligible in comparison, while the background turbulence driven from stellar feedback operates on much larger time and length scales from those of an individual C-shock.

A possible consequence of the turbulence driven by drag-induced instabilities is clump formation. One of the notable examples is the clump formation of dust in the turbulence driven by the streaming instability due to the dust--gas drag in a protoplanetary disk under certain favorable conditions \citep{JY07}. Moreover, we show that the density perturbation dominates over the velocity and magnetic field perturbations in the linear regime of the drag instability, which may hint the dynamical importance of density enhancement.
The observations of nearby molecular clouds with the Herschel Space Observatory suggested that about 70\%-80\% of dense cores lie within filaments \citep{Poly13,Konyves15}. The numerical simulation conducted by \citet{CO12} revealed that a pair of steady C-shocks propagate away from the shocked layer of two colliding flows where prestellar cores and filaments can form \citep{CO14}. The drag instability in a steady C-shock might be potentially capable of core formation lying outside of filaments.  Alternatively, the drag instability is likely to nonlinearly saturate, perhaps at a level as high as the velocity difference across the shock front. This would substantially limit the ability of the instability to drive a subsequent gravitational instability beyond what has already been produced by the jump in density across the shock. In any case,
whether the nonlinear outcome of the density enhancement by the drag instability in a C-shock can facilitate the clump/core formation is beyond the reach of the linear analysis. Given the small  longitudinal  wavelength of the drag instability within a C-shock, nonideal MHD simulations with high resolutions are required to explore this possibility \citep[see the discussion section in][]{GC20}.

Apart from the steady C-shock, a transient C-shock appearing in the shocked layer of two colliding flows has been modeled to be a promising site for the major formation of cores and filamentary structures \citep{NL08,CO12,CO14}. We restrict ourselves to the drag instability in steady C-shocks because the background state has settled to an equilibrium state and thus provides an easy test bed for demonstrating the existence of the drag instability in a linear analysis. Nevertheless,
the ion-neutral drift is expected to be extremely fast in a transient C-shock, perhaps favoring the occurrence of the drag instability \citep{Gu04}. Given the time-dependent nature of a transient C-shock, setting up an appropriate background state for a perturbation theory is expected to be challenging. 
%We leave this intriguing task in a future work.

Undoubtedly, the aforementioned issues and possible implications that we have discussed related to the drag instability are highly speculative and intriguing, requiring prudent and elaborate studies for further investigation. After all, the interplay between ambipolar diffusion, turbulence, shocks, gravity, etc. for the dynamical processes of star formation has been a complex and broad topic.
From a theoretical perspective, the 2D linear analysis conducted in this work based on the 1D analysis presented in \citet{Gu04} and \citet{GC20}  would advance our understanding of the instabilities of astrophysical plasma in general. In practice, our framework
provides the basic properties of the drag instability to be studied in future nonideal MHD simulations for the confirmation of their existence and for understanding the nonlinear outcome of the linear instability in C-shocks.

\acknowledgments
We are grateful to Che-Yu Chen, Min-Kai Lin, Hau-Yu Baobab Liu, and  Chien-Chang Yen for useful discussions. We would also like to thank the referee for helpful comments that greatly improved the manuscript, especially the contents of the discussion section.
This work has been supported by the Ministry of Science and Technology in Taiwan through the grant MOST 109-2112-M001-052.


\begin{thebibliography}{}
\bibitem[Andr\'e et al.(2014)]{Andre14} Andr\'e, P., Di Francesco, J., Ward-Thompson, D., Inutsuka, S. -I., Pudritz, R. E., Pineda, J. E. 2014, Protostars and Planets VI, Henrik Beuther, Ralf S. Klessen, Cornelis P. Dullemond, and Thomas Henning (eds.), University of Arizona Press, Tucson, 914 pp., p.27-51
\bibitem[Ballesteros-Paredes et al.(2007)]{BP07} Ballesteros-Paredes, J., Klessen, R.~S., Mac Low, M.-M., et al.\ 2007, Protostars and Planets V, B. Reipurth, D. Jewitt, and K. Keil (eds.), University of Arizona Press, Tucson, 951 pp., p.63-80
\bibitem[Blitz et al.(2007)]{Blitz} Blitz, L., Fukui, Y., Kawamura, A., Leroy, A., Mizuno, N., \& Rosolowsky, E.,
2007, Protostars and Planets V. Univ. Arizona Press, Tucson, AZ, p. 81
\bibitem[Chen \& Ostriker(2012)]{CO12} Chen, C.-Y, \& Ostriker, E. 2012, \apj, 744, 124
\bibitem[Chen \& Ostriker(2014)]{CO14} Chen, C.-Y, \& Ostriker, E. 2014, \apj, 785, 69
2, \apj, 567, 947
\bibitem[Dalgarno(2006)]{Dalgarno06} Dalgarno, A.\ 2006, Proceedings of the National Academy of Science, 103, 12269
\bibitem[Draine(1980)]{Draine80} Draine, B. T. 1980, \apj, 241, 1021
\bibitem[Draine \& McKee(1993)]{DM93} Draine, B. T., \& McKee, C. F. 1993, ARA\&A, 31, 373
\bibitem[Draine et al(1983)]{Draine} Draine, B. T., Roberge, W. G., \& Dalgarno, A. 1983, \apj, 264, 485
\bibitem[Elmegreen \& Scalo(2004)]{ES04} Elmegreen, B. G., \& Scalo, J. 2004, Annu. Rev. Astron. Astrophys., 42, 211
\bibitem[Engargiola et al.(2003)]{Engargiola} Engargiola, G., Plambeck, R. L., Rosolowsky, E., \& Blitz, L., 2003, \apjs, 149, 343
\bibitem[Falle et al.(2009)]{Falle} Falle, S.A.E.G., Hartquist, T.W., van Loo, S. In: Pogorlov, N.V., Audit, E., Colella, P., Zank, G.P. 2009, (eds.) Numerical Modeling of Space Plasma Flows, p. 80. Astronomical Society of the Pacific
\bibitem[Flower \& Pineau Des For{\^e}ts(1998)]{Flower98} Flower, D.~R., \& Pineau Des For{\^e}ts, G.\ 1998, \mnras, 297, 1182
\bibitem[Flower \& Pineau Des For{\^e}ts(2010)]{Flower10} Flower, D.~R., \& Pineau Des For{\^e}ts, G.\ 2010, \mnras, 406, 1745
\bibitem[Fukui \& Kawamura(2010)]{Fukui10} Fukui, Y., \& Kawamura, A.\ 2010, \araa, 48, 547
\bibitem[Girichidis et al.(2020)]{Girichidis20}  Girichidis, P., Offner, S.S.R., Kritsuk, A.G. et al. 2020, Space Sci Rev, 216, 68
\bibitem[GC20()]{GC20} Gu, P.-G., \& Chen, C.-Y. 2020, \apj, 898, 67 (GC20)
\bibitem[Gu et al.(2004)]{Gu04} Gu, P.-G., Lin, D. N. C., \& Vishniac, E. T., Astrophysics \& Space Science, 292, 261
\bibitem[Gusdorf et al.(2008)]{Gusdorf08} Gusdorf, A., Cabrit, S., Flower, D.~R., et al.\ 2008, \aap, 482, 809
\bibitem[Hennebelle \& Falgarone(2012)]{HF12}  Hennebelle, P., \& Falgarone, E. 2012, Astron. Astrophys. Rev., 20, 55
\bibitem[Hennebelle \& Inutsuka(2019)]{HI19} Hennebelle, P., \& Inutsuka, S.-I. 2019, Frontiers in Astronomy and Space Sciences, 6, 5
\bibitem[Hezareh et al.(2010)]{Hezareh10} Hezareh, T., Houde, M., McCoey, C., et al.\ 2010, \apj, 720, 603
\bibitem[Hezareh et al.(2014)]{Hezareh14} Hezareh, T., Csengeri, T., Houde, M., et al.\ 2014, \mnras, 438, 663
\bibitem[Indriolo \& McCall(2012)]{Indriolo12} Indriolo, N., \& McCall, B.~J.\ 2012, \apj, 745, 91
\bibitem[Jeffreson \& Kruijssen(2018)]{JK18} Jeffreson, S. M. R. \& Kruijssen, J. M. D. 2018, \mnras, 476, 3688
\bibitem[Johansen \& Youdin(2007)]{JY07} Johansen, A., \& Youdin, A. 2007 \apj, 662, 627
\bibitem[Kawamura et al.(2009)]{Kawamura} Kawamura, A. et al., 2009, \apjs, 184, 1
\bibitem[Kennicutt \& Evans(2012)]{KE12} Kennicutt, R. C., \& Evans, N. J. 2012, Annu. Rev. Astron. Astrophys. 50, 531
\bibitem[K\"onyves et al.(2015)]{Konyves15} K\"onyves, V., Andr\'e, P., Men\'shchikov, A., Palmeirim, P., Arzoumanian, D.,
Schneider, N., et al. 2015, \aap, 584, 91
%\bibitem[Kulsrud \& Pearce(1969)]{KP69} Kulsrud, R., \& Pearce, W. L. 1969, \apj, 156, 445 
\bibitem[Lehmann \& Wardle(2016)]{LehmannWardle16} Lehmann, A., \& Wardle, M.\ 2016, \mnras, 455, 2066
\bibitem[Li \& Nakamura(2004)]{LN04} Li, Z.-Y., \& Nakamura, F. 2004, \apjl, 609. 83
\bibitem[Li et al.(2014)]{LiHB14} Li, H.-B., Goodman, A., Sridharan, T.~K., et al.\ 2014, Protostars and Planets VI, 101
\bibitem[Li \& Houde(2008)]{LiHoude08} Li, H.-B., \& Houde, M.\ 2008, \apj, 677, 1151
\bibitem[Mac Low(1999)]{MacLow99} Mac Low, M.-M.  1999, \apj, 524, 169
%\bibitem[Mac Low \& Smith(1997)]{MacLow97} Mac Low, M.-M., \& Smith, M. D. 1997, \apj, 491, 596
\bibitem[Mac Low et al.(1998)]{MacLow98} Mac Low, M.-M., Klessen, R. S., Burkert A. et al., 1998, Phys. Rev. Lett. 80, 2754
\bibitem[McKee et al.(2010)]{McKee10} McKee, C. F., Li, P. S., \& Klein, R. I. 2010, \apj, 720, 1612
\bibitem[Meidt et al.(2015)]{Meidt} Meidt, S. E. et al., 2015, \apj, 806, 72
\bibitem[Mestel \& Spitzer(1956)]{MS1956} Mestel, L., \& Spitzer, L.\ 1956, \mnras, 116, 503
\bibitem[Miura et al.(2012)]{Miura} Miura, R. E. et al., 2012, \apj, 761, 37
%\bibitem[Mouschovias(1979)]{Mouschovias79} Mouschovias, T.~C.\ 1979, \apj, 228, 475
\bibitem[Murray(2011)]{Murray} Murray, N., 2011, \apj, 729, 133
\bibitem[Nakamura \& Li(2005)]{NL05} Nakamura, F., \& Li, Z.-Y. 2005, \apj, 631, 411
\bibitem[Nakamura \& Li(2008)]{NL08} Nakamura, F., \& Li, Z.-Y. 2008, \apj, 687, 354
%\bibitem[Pineau des For{\^e}ts et al.(1997)]{Pineau97} Pineau des Fore{\^e}ts, G., Flower, D.~R., \& Chieze, J.-P.\ 1997, in IAU Symp. 182, Herbig-Haro Flows and the Birth of Stars, ed. B. Reipurth \& C. Bertout (Dordrecht: Kluwer), 199
\bibitem[Polychroni et al.(2013)]{Poly13} Polychroni, D., Schisano, E., Elia, D., Roy, A., Molinari, S., Martin, P., et al. 2013, \apjl,777, 33 
\bibitem[Shu(1992)]{Shu} Shu, F. H. 1992, in Physics of Astrophysics, Vol. II, ed. F. H. Shu (Mill Valley, CA: Univ. Science Books)
\bibitem[Smith \& Mac Low(1997)]{SmithML97} Smith, M.~D., \& Mac Low, M.-M.\ 1997, \aap, 326, 801
\bibitem[Spitzer(1956)]{Spitzer1956} Spitzer, L.\ 1956, Physics of Fully Ionized Gases, New York: Interscience Publishers
\bibitem[Stone(1997)]{Stone} Stone, J. M. 1997, \apj, 487, 271
\bibitem[Stone et al.(1998)]{Stone98} Stone, J. M., Ostriker, E. C., Gammie, C. F. 1998, \apjl  508, 99
\bibitem[Tang et al.(2018)]{Tang18} Tang, K.~S., Li, H.-B., \& Lee, W.-K.\ 2018, \apj, 862, 42
\bibitem[Tielens(2005)]{Tielens05} Tielens, A.~G.~G.~M.\ 2005, The Physics and Chemistry of the Interstellar Medium, Cambridge, UK: Cambridge University Press
\bibitem[Valdivia et al.(2017)]{Valdivia17} Valdivia, V., Godard, B., Hennebelle, P., et al.\ 2017, \aap, 600, A114
\bibitem[Wardle(1990)]{Wardle} Wardle, M. 1990, \apj, 246, 98
\bibitem[Wardle(1991)]{Wardle1991} Wardle, M.\ 1991, \mnras, 251, 119
\bibitem[Wardle(1998)]{Wardle1998} Wardle, M. 1998, \mnras, 298, 507
\bibitem[Xu \& Li(2016)]{XuLi16} Xu, D., \& Li, D.\ 2016, \apj, 833, 90
\bibitem[Zweibel(2015)]{Zweibel} Zweibel, E. 2015, Ambipolar Diffusion, in Magnetic Fields in Diffuse Media, ed. Alexander Lazarian, Elisabete M. de Gouveia Dal PinoClaudio Melioli, 407, 285
\end{thebibliography}
\end{document}